\documentclass[12pt]{article}
\usepackage{titling}
\usepackage{amsmath}
\usepackage{slashed}
\usepackage{amssymb}
\usepackage{epsfig}
\usepackage{graphicx}
\usepackage{multirow}
\usepackage{color}
\usepackage[normal]{subfigure}
\usepackage{rotating}
\usepackage{hyperref}
\usepackage[margin=0.9in]{geometry}
\usepackage[table]{xcolor}
\usepackage{enumitem}
\usepackage[utf8x]{inputenc}
\usepackage[compress,numbers,sort]{natbib}
\usepackage{authblk}
\usepackage{colortbl}
\usepackage{pdflscape}
\usepackage{color}

\def\eq#1{{Eq.~(\ref{#1})}}
\def\eqs#1#2{{Eqs.~(\ref{#1})--(\ref{#2})}}
\def\fig#1{{Fig.~\ref{#1}}}

\def\Table#1{{Table~\ref{#1}}}

\def\sect#1{{Sect.~\ref{#1}}}

\def\app#1{{Appendix~\ref{#1}}}

\def\vev#1{\left\langle #1\right\rangle}
\def\abs#1{\left| #1\right|}

\def\Im{\mbox{Im}\,}
\def\Re{\mbox{Re}\,}

\newcommand{\be}{\begin{eqnarray}}
\newcommand{\ee}{\end{eqnarray}}

\newcommand{\bee}{\begin{eqnarray}}
\newcommand{\eee}{\end{eqnarray}}
\newcommand{\beeq}{\begin{equation}}
\newcommand{\eeeq}{\end{equation}}

\newcommand{\vs}{v_{\scriptscriptstyle S}}
\newcommand{\vh}{v_{\scriptscriptstyle H}}
\newcommand{\vt}{v_{\scriptscriptstyle T}}
\newcommand{\lambdah}{\lambda_1}

\DeclareRobustCommand{\Fig}[1]{Fig.~\ref{#1}}

\definecolor{nicered}{rgb}{0.7,0.1,0.1}
\definecolor{nicegreen}{rgb}{0.1,0.5,0.1}
\definecolor{red}{rgb}{1.0, 0, 0}

\newcommand{\beq}{\begin{equation}}
\newcommand{\eeq}{\end{equation}}
\newcommand{\bea}{\begin{eqnarray}}
\newcommand{\eea}{\end{eqnarray}}

\title{\bf{Maxi-sizing the trilinear Higgs self-coupling: how large could it be?}}

\author{Luca Di Luzio\thanks{luca.di-luzio@durham.ac.uk}}
\author{Ramona Gr\"ober\thanks{ramona.groeber@durham.ac.uk}}
\author{Michael Spannowsky\thanks{michael.spannowsky@durham.ac.uk}}

\affil{\emph{\normalsize Institute for Particle Physics Phenomenology, Department of Physics, 
Durham University, Durham DH1 3LE, United Kingdom}}

\date{}

\begin{document} 

\maketitle
\begin{abstract}
\normalsize
In order to answer the question on how much the trilinear Higgs self-coupling 
could deviate from its Standard Model value in weakly coupled models, we study both theoretical and phenomenological constraints. 
As a first step, we discuss this question by modifying the Standard Model using effective operators. 
Considering constraints from vacuum stability and perturbativity, we show that only the latter can be 
reliably assessed in a model-independent way. We then focus on UV models 
which receive constraints from 
Higgs coupling measurements, electroweak precision tests, vacuum stability and perturbativity. 
We find that the interplay of current measurements with 
perturbativity already exclude self-coupling modifications above 
a factor of few with respect to the Standard Model value.
\end{abstract}

\clearpage

\tableofcontents

\clearpage

\section{Introduction}
\label{sec:intro}

The recent discovery of the Higgs boson at the Large Hadron Collider (LHC) 
\cite{Aad:2012tfa,Chatrchyan:2012xdj} marks a milestone event for high-energy physics.
 Yet, 
 the Higgs boson is only a remnant of the underlying mechanism of spontaneous electroweak (EW) 
 symmetry breaking, the so-called Brout-Englert-Higgs mechanism \cite{Englert:1964et,Higgs:1964pj}. 
  In order to improve our understanding of the dynamics initiating EW symmetry breaking, 
  a key ingredient is the 
  global structure of the scalar potential that triggers the spontaneous breaking of 
  $SU(2)_L \times U(1)_Y \to U(1)_\mathrm{QED}$.
While the ongoing LHC program, focusing on precise measurements of Higgs and gauge boson
 masses and couplings, will continue to improve our understanding of the potential's local
  structure in the vicinity of the EW minimum, information on the shape of the vacuum in a
   model-independent way is experimentally very difficult to obtain.\footnote{The energy 
   scale of non-perturbative phenomena, e.g.~the mass of the $SU(2)_L$ sphalerons 
   \cite{Klinkhamer:1984di}, could potentially allow to probe the scalar potential
    away from the EW minimum \cite{Spannowsky:2016ile}.} 

However, if one specifies the degrees of freedom and interactions in the scalar sector,
one can calculate the form of the scalar potential. After EW symmetry breaking such potential gives
rise to multi-scalar interactions, i.e.~at lowest order cubic and quartic Higgs
self-interactions. While the former can be probed directly in searches for 
multi-Higgs final states \cite{ATLAS-CONF-2016-004,Aaboud:2016xco,ATLAS-CONF-2016-071,
CMS-PAS-HIG-16-026,CMS-PAS-HIG-17-002,CMS-PAS-HIG-17-006, Djouadi:1999gv, Djouadi:1999rca, Baur:2002qd, Baur:2003gp,
Dolan:2012rv,Baglio:2012np,Barr:2013tda,Dolan:2013rja,Papaefstathiou:2012qe,Goertz:2013kp,
deLima:2014dta,Englert:2014uqa,Liu:2014rva,Goertz:2014qta,Azatov:2015oxa,Carvalho:2015ttv,Behr:2015oqq},
indirectly via their effect on precision observables \cite{Degrassi:2017ucl,Kribs:2017znd}
or loop corrections to single Higgs production \cite{McCullough:2013rea, Gorbahn:2016uoy,
Degrassi:2016wml,Bizon:2016wgr,DiVita:2017eyz},
the latter are inaccessible at the LHC or a future linear collider \cite{Plehn:2005nk,Binoth:2006ym, Battaglia:2004mw}. 
Thus, to obtain a glimpse at the shape of the scalar potential we have to focus on the cubic scalar self-coupling.

If new light degrees of freedom contribute to the Higgs potential, they typically dominate the multi-Higgs phenomenology.
On the other hand, if new degrees of freedom are heavy, it is widely argued that the Effective Field Theory (EFT)
approach is most suitable to study deformations of the Standard Model (SM) Higgs potential in a rather model-independent 
and predictive way. Thus, in the latter case, where we assume that no light states below the cutoff
scale $\Lambda \gg v \equiv 246$ GeV exist, it is tempting to introduce an operator $|H|^6$ (where $H$ denotes the usual Higgs 
doublet)
and connect the (global) properties of the vacuum, e.g.~whether the EW minimum is a local or 
global one, with the cubic Higgs self-coupling. In particular, one could consider using vacuum stability
arguments to infer model-independent bounds on the triple Higgs coupling.

In this work, we show that this approach is flawed. In particular, there can be two kinds of instabilities 
corresponding to the possible emergence of new minima either at large field values 
$v \ll \bar{h} \lesssim \Lambda$ or at $\bar{h}=0$ 
(where $\bar{h}$ denotes the background field of the effective Higgs potential, whose minimum determines the ground state 
of the theory). The former, is shown to be spurious 
since the very expansion of the scalar potential in powers of $\bar h/\Lambda$ in the vicinity of an instability 
leads to the breakdown of the EFT expansion \cite{Burgess:2001tj}. 
In \sect{sec:EFT} we explicitly show that a weakly coupled toy
 model can feature an absolutely stable vacuum in the full theory, 
 while obtaining a spurious instability in the EFT limit. 
Similarly, the second type of instability, 
due to the emergence of a new minimum in $\bar{h}=0$, is also shown to be 
not under control when including only the lowest terms in the EFT expansion. 

On the other hand, allowing for too large Higgs self-couplings (either trilinear or quadrilinear ones) 
raises the question of the validity of perturbative methods. 
When tree-level scattering amplitudes violate unitarity, large 
higher-order corrections are necessary to restore unitarity, thus leading to the breakdown of the 
perturbative expansion. This argument has been
employed in the past to set theoretical bounds on couplings and scales.
The most famous example
is the scattering of longitudinal vector bosons, which has been used to set a theoretical 
limit on the Higgs boson mass by
performing a partial wave analysis \cite{Cornwall:1974km,Lee:1977yc}.
We apply this method in Sect.~\ref{sectperthhh}
in order to set a bound on Higgs self-couplings 
by considering the $hh\to hh$ scattering.
In addition, we show that the requirement that the loop-corrected Higgs scalar vertices are 
smaller than their tree-level values gives a very similar theoretical bound on Higgs self-couplings.

Given the apparent limitations of the EFT framework in setting bounds beyond perturbativity, 
we focus on UV complete scenarios from \sect{UVmodels} onwards 
to investigate the question of the maximally allowed triple Higgs coupling. 
We consider for simplicity only weakly coupled models, as they retain a higher degree of predictivity 
and we have full control of the theory. Particularly large deviations are 
expected in scenarios where the SM is augmented by extra scalars.
We focus on new scalars $\Phi$ which can couple via a tadpole operator of the type 
$\mathcal{O}_{\Phi} = \Phi f(H)$, where $f(H)$ is a string of Higgs fields 
(or their charge conjugates). In \sect{UVmodels} we argue that such 
couplings potentially give the largest contributions to the Higgs self-coupling 
and classify all the possible representations of $\Phi$ that lead 
to such interactions. As a result of the presence of the new scalars, 
the vacuum structure of the scalar potential is more contrived and
it becomes challenging to establish a direct relation between 
Higgs self-coupling deviations and the stability of the EW vacuum. 
Still, parts of the parameter space can be excluded by requiring the vacuum to be (meta)stable. 
In addition, we take into account phenomenological limits from Higgs coupling measurements and EW 
precision tests. Together with a perturbativity requirement for the parameters of the extended scalar potential,
we find that maximal deviations up to few times the SM trilinear Higgs self-coupling are still feasible.

Looking beyond tree level, we investigate
loop-induced modifications in \sect{looptrilinear}. 
While such contributions are expected to be smaller, 
they are of particular interested as they are induced by a
plethora of new physics models. 
We discuss here the case of fermionic loops, since in such a case one can
regain a direct correlation between the triple Higgs coupling and the 
stability of the EW vacuum. We comment on this relation, explicitly studying 
the case of low-scale seesaw models, which are largely unconstrained by other Higgs couplings' measurements. 
Finally, in \sect{sec:conclusions} we present our conclusions.

\section{Theoretical constraints on Higgs self-couplings}
\label{sec:EFT}

Let us parametrize the Higgs potential in the SM broken phase 
as
\beq 
\label{V3heff}
V (h) = \frac{1}{2} m_h^2 h^2 + \frac{1}{3!} \lambda_{hhh} h^3 
+ \frac{1}{4!} \lambda_{hhhh} h^4  \, , 
\eeq
where $h$ denotes the CP-even neutral components 
of the Higgs doublet, i.e.~$H = \frac{1}{\sqrt{2}} (0, v + h)^T$ in the unitary gauge, 
and $\lambda_{hhh}$ ($\lambda_{hhhh}$) is the modified trilinear (quadrilinear) Higgs self-coupling. 
In the SM we have 
\beq 
\label{deflamhhhSM}
\lambda_{hhh}^{\text{SM}} = \frac{3 m^2_h}{v} \simeq 190 \ \text{GeV} \qquad \text{and} \qquad
\lambda_{hhhh}^{\text{SM}} = \frac{3 m^2_h}{v^2} \simeq 0.77 \, .
\eeq
The question we want to address is whether there exist some model-independent
bounds on the value of the Higgs self-couplings. 
To this end, we will consider two classes of theoretical constraints which are vacuum stability and perturbativity. While the latter is, strictly speaking, not a bound, it is still 
interesting given our limitations in using \eq{V3heff} beyond perturbation theory. 
In \sect{sectperthhh} we will provide a simple perturbativity criterium which can be applied to the potential of \eq{V3heff}. 
On the other hand, in order to formulate the question of vacuum stability in a gauge invariant way we will add an operator 
$\frac{c_6}{v^2} \abs{H}^6$ to the SM Lagrangian and study the vacuum structure of the theory. 
Would then be possible to set model-independent bounds on the Wilson coefficient $c_6$ from the 
requirement that the EW vacuum is absolutely stable or long-lived enough?  
As we are going to see, the answer to the this question is in general negative, 
requiring a careful analysis of the range of applicability 
of the EFT.

\subsection{EW symmetry breaking with $d=6$ operators}
\label{EWSBdim6}

We start by reviewing EW symmetry breaking in the SM augmented by the operator $\abs{H}^6$ (see e.g.~\cite{Barger:2003rs}). 
The truncated potential reads
\beq
\label{V6}
V^{(6)} (H) = -\mu^2 \abs{H}^2 + \lambda \abs{H}^4 + \frac{c_6}{v^2} \abs{H}^6 \, ,
\eeq
where the normalization of the $d=6$ operator is given in terms of $v = (\sqrt{2} G_\mu)^{-1/2} \simeq 246$ GeV. 
Note that $c_6 = \bar{c}_6 \lambda$ in the notation of Ref.~\cite{Contino:2013kra}.
In the following, we will focus on weakly coupled regimes, 
where $c_6$ is at most of $\mathcal{O}\left(v^2 / \Lambda^2\right)$
and $\Lambda$ is the cutoff of the 
EFT.\footnote{By naive dimensional analysis the scaling of $c_6$ is $g_*^4 v^2/\Lambda^2$, where 
$g_*$ denotes a generic coupling which can range up to $4 \pi$ in strongly-coupled theories 
(see e.g.~\cite{Pomarol:2014dya}). However, in theories where the Higgs mass is protected by an additional symmetry, like e.g.~in 
composite Higgs models, the scaling of the coefficient $c_6$ is expected to be 
$c_6 \sim \lambda g_*^2 v^2/\Lambda^2=\lambda v^2/f^2$, with 
$1/f \equiv g_*/\Lambda$ \cite{Giudice:2007fh,Contino:2013kra}. 
Hence, also in this case values of $c_6 \sim 1$ lead to the 
breakdown of the EFT expansion.}

In order to minimize the potential, we project the Higgs doublet on its background real component, 
$H \rightarrow \frac{1}{\sqrt{2}} \bar h$.
From the equation 
\beq
\label{stateq}
V^{(6)'} (\bar h) = \left( - \mu^2 + \lambda \bar h^2 + \frac{3 c_6}{4 v^2} \bar h^4 \right) \bar h = 0 \, , 
\eeq 
we find three possible stationary points: $\bar h = 0, \, v_+, \, v_-$, where  
\beq 
\label{vpmgeneral}
v^2_{\pm} = \frac{2 v^2}{3 c_6} \left( -\lambda \pm \abs{\lambda} \sqrt{1 + \frac{3 c_6 \mu^2}{\lambda^2 v^2} } \right) 
= \left( \pm \abs{\lambda} -\lambda  \right)  \frac{2 v^2}{3 c_6} \pm \frac{\mu^2}{\abs{\lambda}} 
\mp \frac{3 c_6 \mu^4}{4 \abs{\lambda}^3 v^2} + \mathcal{O}(c_6^2)
\, ,
\eeq 
and in the last step we expanded for $c_6 \ll 1$. 
The nature of the stationary points (whether they correspond to maxima or minima) 
depends on the second derivative of the potential
\beq
\label{V6second}
V^{(6)''} (\bar h) = - \mu^2 + 3 \lambda \bar h^2 + \frac{15 c_6}{4 v^2}  \bar h^4 \, . 
\eeq 
Considering the possible signs of the potential parameters in \eq{V6} we have in total $2^3=8$ combinations, 
out of which only 4 lead to a phenomenologically viable (i.e.~$\bar h \neq 0$) EW minimum:
\begin{enumerate}
\item $\mu^2 > 0$, $\lambda > 0$, $c_6 > 0$: In this case \eq{vpmgeneral} yields 
(at the next-to-leading order in the $c_6$ expansion)
\begin{align}
\label{vpcase1}
v^2_+ & \simeq \frac{\mu^2}{\lambda} \left( 1 - \frac{3 c_6 \mu^2}{4 \lambda^2 v^2} \right) \, , \\
\label{vmcase2}
v^2_- & \simeq - \frac{4 \lambda v^2}{3 c_6} \left( 1 + \frac{3 c_6 \mu^2}{4 \lambda^2 v^2} \right) \, .   
\end{align} 
As $c_6 > 0$, only $v_+$ is a stationary point and from \eq{V6second} we find 
\begin{align}
V^{(6)''} (0) &= - \mu^2 < 0 \, , \\
V^{(6)''} (v_+) &\simeq 2 \mu^2 \left( 1 + \frac{3 c_6 \mu^2}{4 \lambda^2 v^2} \right) > 0 \, . 
\end{align}
Hence, $\bar h = 0$ is a maximum, while $\bar h = v_+$ can be identified with the EW minimum $v$. 
Note that in the $c_6 \to 0$ limit we recover the SM result. 

\item $\mu^2 > 0$, $\lambda > 0$, $c_6 < 0$: In addition to $\bar h = 0$ and $v_+$, as before, 
we have a third stationary point $v_-$, as now $c_6 < 0$ (cf.~\eq{vmcase2}). 
The latter corresponds to a maximum, as implied by
\beq
V^{(6)''} (v_-) \simeq \frac{8 \lambda^2 v^2}{3 c_6} \left( 1 + \frac{9 c_6 \mu^2}{4 \lambda^2 v^2} \right) < 0 \, . 
\eeq
The potential, which is sketched in the left panel of \fig{fig:Instability_H}, 
features an instability at large field values $\bar h \gtrsim v_- \sim \sqrt{\lambda} \Lambda$ (where we used $c_6 \sim v^2 / \Lambda^2$). 
The instability looks however specious, because it is close to the cutoff of the EFT. 
As in the previous case, for $c_6 \to 0$ we recover the SM since the position of the second maximum is pushed to infinity.   
\item $\mu^2 < 0$, $\lambda < 0$, $c_6 > 0$: Substituting in \eq{vpmgeneral} we get 
\begin{align}
\label{vpcase3}
v^2_+ & \simeq \frac{4 \abs{\lambda} v^2}{3 c_6} \left( 1 + \frac{3 c_6 \mu^2}{4 \lambda^2 v^2} \right) \, , \\
\label{vmcase3}
v^2_- & \simeq - \frac{\mu^2}{\abs{\lambda}} \left( 1 - \frac{3 c_6 \mu^2}{4 \lambda^2 v^2} \right) \, ,   
\end{align} 
while the second derivatives of the potential read
\begin{align}
\label{Vp0case3}
V^{(6)''} (0) &= - \mu^2 > 0 \, , \\
\label{Vpvpcase3}
V^{(6)''} (v_+) &\simeq \frac{8 \lambda^2 v^2}{3 c_6} \left( 1 + \frac{9 c_6 \mu^2}{4 \lambda^2 v^2} \right) > 0 \, , \\
\label{Vpvmcase3}
V^{(6)''} (v_-) &\simeq 2 \mu^2 \left( 1 + \frac{3 c_6 \mu^2}{4 \lambda^2 v^2} \right) < 0 \, .
\end{align}
Thus $\bar h = 0$ and $v_+$ are 
minima, while $v_-$ is a maximum. Note that the potential gets flipped when compared to 
that of case 2.~(cf.~solid curve 
in the right panel of \fig{fig:Instability_H}). This time, however, we must identify the EW minimum $v$ 
with $v_+ \sim \sqrt{\abs{\lambda}} \Lambda$ (where we used $c_6 \sim v^2 / \Lambda^2$), 
which means that the EW vacuum expectation value (vev) is generated by the physics at the cutoff scale. 
This corresponds to a non-decoupling EFT, since in the $c_6 \to 0$ limit 
the EW minimum is pushed to infinity and we do not re-obtain the SM. 

\item $\mu^2 > 0$, $\lambda < 0$, $c_6 > 0$: This case is similar to the previous one, with the difference that 
$\bar h = 0$ is a maximum (cf.~\eq{Vp0case3}), 
the maximum in $v_-$ disappears (cf.~\eq{vmcase3}), while the $\Lambda$ dominated EW minimum 
remains in $v_+$ (cf.~\eq{Vpvpcase3}). Also in this case the limit $c_6 \to 0$ does not reproduce the SM. 

\end{enumerate}

\subsection{Vacuum instabilities}
\label{EFTvac}

\begin{figure}[h!]
\centering
\includegraphics[angle=0,width=7.cm]{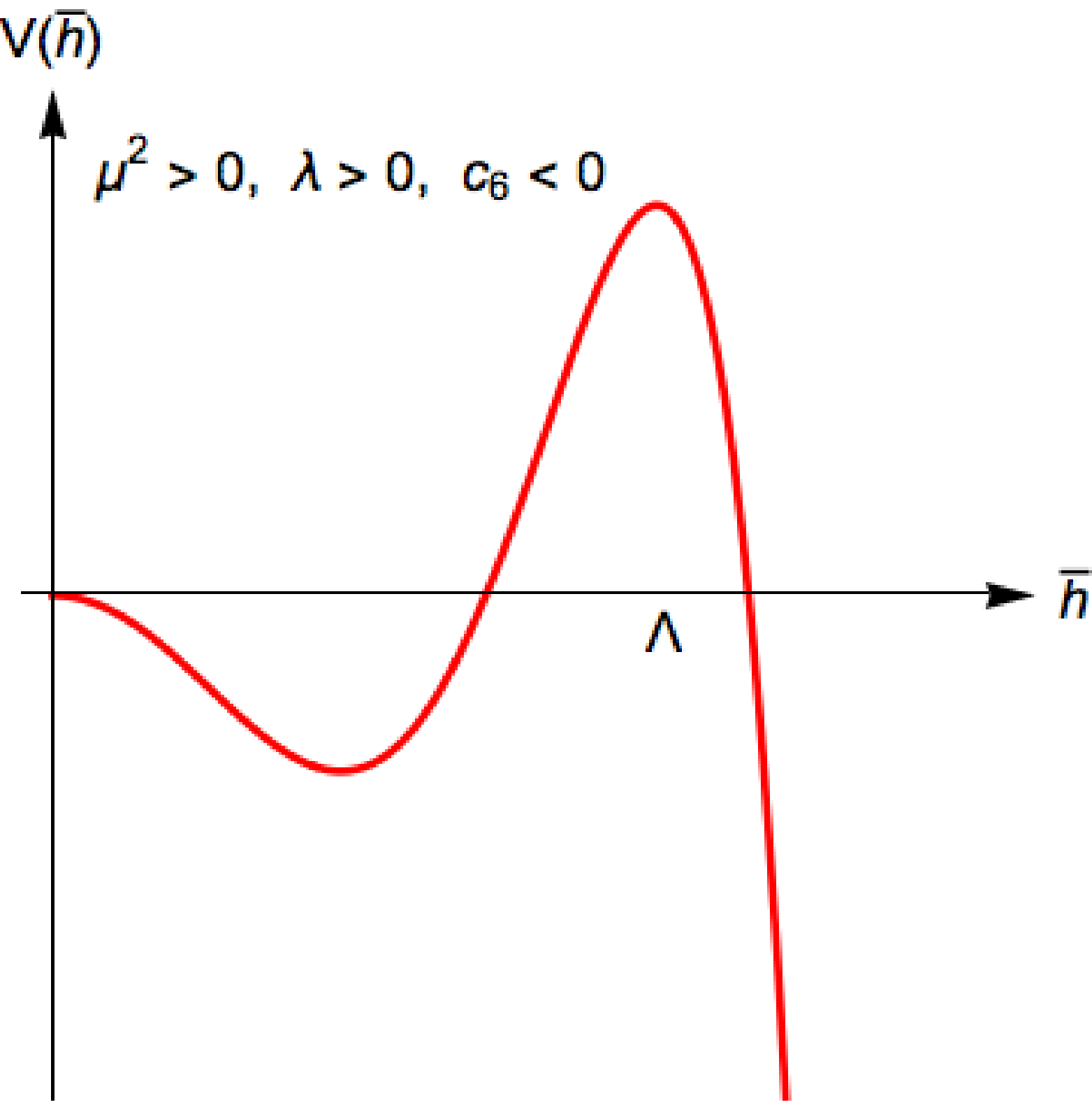} \ \ \ \ \
\includegraphics[angle=0,width=7.cm]{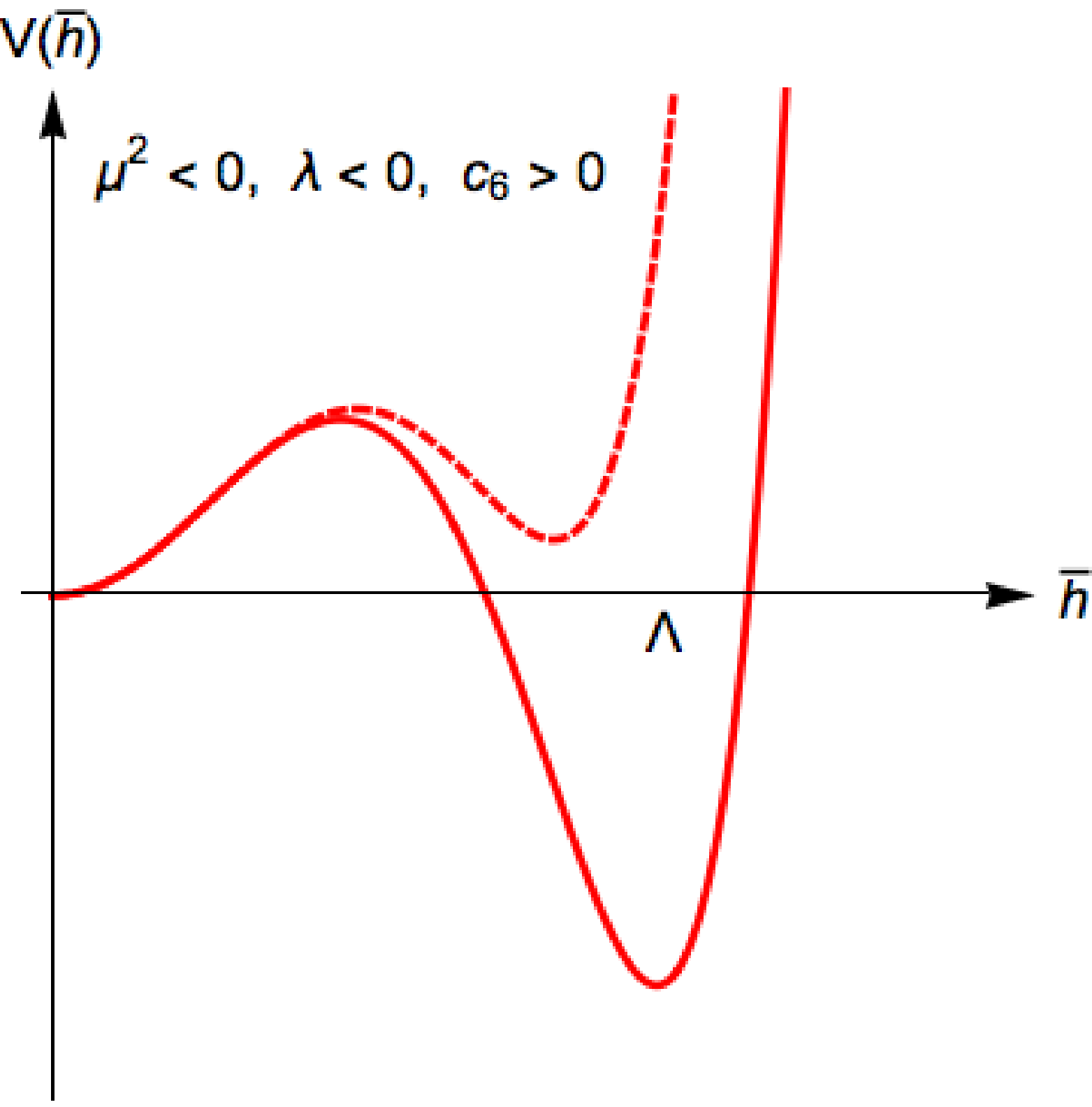}
\caption{\label{fig:Instability_H} 
The two kind of instabilities triggered by a sizable $c_6$. 
{\it Left:} A negative $c_6$ is responsible for a large-field-value instability close to the scale $\Lambda$. 
{\it Right:} The EW minimum is generated by the physics at the cutoff scale $\Lambda$. 
For large enough $c_6>0$, the absolute minimum is in $\bar h=0$ (dashed line), and the 
EW vacuum gets destabilized. 
}
\end{figure}

There are essentially two types of instabilities associated with the presence of the coupling $c_6$: 
the most obvious one, at large field values, is triggered by a negative $c_6$ (case 2~in \sect{EWSBdim6}), 
while the other one has to do with the destabilization of the EW ground state against the minimum 
in $\bar h = 0$ (case 3~in \sect{EWSBdim6}), which happens for large, positive, values of $c_6$ 
(dashed curve in the right plot of \fig{fig:Instability_H}). 
This might suggest that there is a lower and upper bound on $c_6$ 
by the requirement that the EW minimum 
is the absolute one.  
However, we are going to argue that there is no such a model-independent bound within a generic 
EFT.
Let us discuss in turn the two kind of instabilities. 

\subsubsection{Large-field-value instability: $\bar h \lesssim \Lambda$}

The main observation here is that 
the very expansion of the scalar potential in powers of $\bar h/\Lambda$ in the vicinity of an instability 
leads to the breakdown of the EFT expansion \cite{Burgess:2001tj}.\footnote{This instability
was discussed in a slightly different context in 
Ref.~\cite{Burgess:2001tj}. There it was shown that the effect of an $\abs{H}^6$ operator on the vacuum stability analysis of the SM is 
always small, whenever it can be reliably computed within the EFT.} 
This has to be traced back to the fact that when the scalar potential is close to vanish, 
field configurations $\bar h \sim \Lambda$ do not cost prohibitive energy to excite, 
contrary to the standard case $V(\bar h \sim \Lambda) \sim \Lambda^4$.  

The spurious nature of the $\abs{H}^6$ instability
is clearly exemplified by taking the EFT limit of a 
simple toy model that features, by construction, absolute stability in the full 
theory \cite{Burgess:2001tj}. Let $h$ and $\phi$ be two real scalar fields, whose potential reads 
\beq 
\label{Vhphitoym}
V(h, \phi) = - \frac{1}{2} m^2 h^2 + \frac{1}{4} \lambda h^4 + \frac{1}{2} M^2 \phi^2 
+ \xi h^3 \phi + \kappa h^2 \phi^2 + \frac{1}{4} \lambda' \phi^4 \, . 
\eeq
Let us consider now the limit $M^2 \gg m^2 > 0$. 
The stationary equations can be solved perturbatively for $m^2/M^2 \ll 1$, 
thus yielding
\begin{align}
\label{vevhtoy}
\vev{h} &\simeq \left( \frac{m^2}{\lambda} \right)^{\frac{1}{2}}\, , \\
\label{vevphitoy}
\vev{\phi} &\simeq - \frac{\xi}{M^2} \left( \frac{m^2}{\lambda} \right)^\frac{3}{2} \ll \vev{h} \, ,  
\end{align}
which is a global minimum as long as $M^2 > \frac{9 \xi^2}{2 \lambda^2} m^2$. 
Moreover, a sufficient condition for the potential to be bounded from below is 
\beq 
\label{stabilitytoym}
\kappa > 0 \, , \quad \land \quad \lambda > \frac{\xi^2}{\kappa}  \, , \quad \land \quad \lambda' > 0 \, , 
\eeq 
so by choosing the potential parameters as in \eq{stabilitytoym} it is always possible to ensure that the 
vacuum in \eqs{vevhtoy}{vevphitoy} is absolutely stable.

Now we integrate $\Phi$ out. A standard calculation yields 
\beq 
\label{EFTtoy1}
V_{\text{EFT}}(h) \simeq - \frac{1}{2} m^2 h^2 + \frac{1}{4} \lambda h^4 - \frac{1}{2} \xi^2 \frac{h^6}{M^2 + 2 \kappa h^2} \, . 
\eeq  
As a consequence of \eq{stabilitytoym} the EFT potential in \eq{EFTtoy1} is clearly stable as well. 
On the other hand, by expanding the denominator of the $h^6$ term for $M^2 \gg 2 \kappa h^2$, we get 
\beq 
V_{\text{EFT}}(h) \simeq - \frac{1}{2} m^2 h^2 + \frac{1}{4} \lambda h^4 
- \frac{1}{2} \frac{\xi^2}{M^2} h^6 
+ \frac{\xi^2 \kappa}{M^4} h^8 + \ldots \, . 
\eeq  
Apparently, the $h^6$ operator features an instability, which however is not supported by 
the full renormalizable model in view of the stability conditions in \eq{stabilitytoym}.
The key point is that the spurious instability sourced by the $h^6$ term does not capture the $\kappa$ 
dependence, as the appropriate resummation of the geometric series shows in \eq{EFTtoy1}. 
We hence conclude that it is not possible to set a model-independent bound on 
$c_6$ from the requirement of stability at large field values. 

We finally note that a possible gauge-invariant way to realize the toy model in \eq{Vhphitoym} 
is given by an Higgs doublet $H \sim (1,2,1/2)$ 
(where the quantum numbers in the bracket denote the transformation properties under $SU(3)_c \times SU(2)_L \times U(1)_Y$) 
coupled to an EW quadruplet $\Phi \sim (1,4,-3/2)$
via the scalar potential 
\begin{align}
\label{scalpot1432}
V(H,\Phi) &= - \mu^2_H \abs{H}^2 +  \mu^2_\Phi \abs{\Phi}^2 + \lambda_H \abs{H}^4 
+ \lambda_1 \abs{H}^2 \abs{\Phi}^2 + \lambda_2 H^* H \Phi^* \Phi \nonumber \\
& + \left( \lambda_3 H H H \Phi + \text{h.c.} \right)  
+ \lambda_\Phi \abs{\Phi}^4 
+ \tilde{\lambda}_\Phi \Phi^* \Phi \Phi^* \Phi \, , 
\end{align}
where non-trivial $SU(2)_L$ contractions are left understood. We have checked that the same qualitative conclusions obtained 
within the toy model apply to the more realistic case of \eq{scalpot1432}.

\subsubsection{Low-scale instability: $\bar h = 0$}

In order to study this case it is more convenient 
to trade the parameters $\mu^2$ and $\lambda$ in terms of the EW vev 
$v$ and the physical Higgs mass $m_h$. 
Imposing the existence of the EW minimum $\bar h = v$ from \eq{stateq} and 
expanding over the Higgs field fluctuations $v \to v + h$, one gets
\begin{align}
\label{m2physpaer}
\mu^2 &= \lambda v^2 + \frac{3}{4} c_6 v^2 = \frac{m^2_h}{2} - \frac{3}{4} c_6 v^2 \, , \\
\label{lamphyspaer}
\lambda &= \frac{m_h^2}{2 v^2} - \frac{3}{2} c_6 \, .  
\end{align} 
By substituting $v=246$ GeV and $m_h=125$ GeV in \eqs{m2physpaer}{lamphyspaer}, 
we find $\mu^2 < 0$ and $\lambda < 0$ as long as $c_6 \gtrsim 0.17$. This is precisely the 
situation described in case 3 of \sect{EWSBdim6}. By taking an even larger $c_6$ 
the minimum in $\bar h = 0$ might become the absolute one (cf.~\fig{fig:Instability_H}). 
This happens for (see also \cite{Degrassi:2016wml})
\beq 
V^{(6)} (v) = \frac{c_6 v^4 - m^2_h v^2}{8} > 0 = V^{(6)} (\bar h = 0) \, ,  
\eeq
corresponding to $c_6 \gtrsim 0.26$. 
However, for a weakly coupled theory where $c_6$ scales like $v^2 / \Lambda^2$, 
such value of $c_6$ implies a very low cutoff scale of $\Lambda \lesssim 480$ GeV, thus 
making the application of the EFT questionable. On the other hand, even admitting for a strongly-coupled origin of $c_6$,  
higher-order operators cannot be consistently 
neglected for assessing the global structure of the Higgs potential away from the EW minimum, 
since $\abs{H}^6$ gives access only up to the sixth derivative of the potential on the EW minimum.

\subsection{Perturbativity bounds} 
\label{sectperthhh}

On general grounds, one expects that too large values of the Higgs self-couplings are bounded 
by perturbativity arguments. In the following, we compare two criteria: the former is based on the partial-wave unitarity 
of the Higgs bosons' scattering amplitude, while the latter consists in the requirement that the loop corrections 
to the Higgs self-interaction vertices are smaller than the tree-level ones. Both criteria yield a similar result. 

\subsubsection{Partial-wave unitarity}
\label{partialwave}

\begin{figure}[h!]
\centering
\includegraphics[angle=0,width=15.cm]{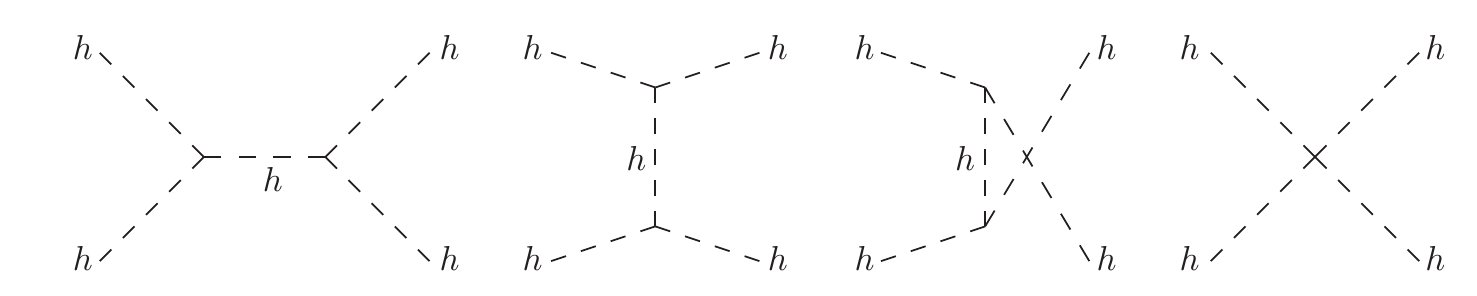} 
\caption{\label{fig:hscattering} 
$hh \to hh$ scattering amplitudes: $s+t+u$ channels $+$ 4-vertex (4$\text{vrtx}$) contributions. 
}
\end{figure}

The $2 \to 2$ Higgs bosons' scattering amplitude 
grows for large values of the Higgs self-couplings, 
eventually leading to unitarity violation and hence to the breakdown 
of the perturbative expansion.\footnote{A similar approach was used in order to set constraints on the size of 
MSSM trilinear couplings (see e.g.~\cite{Schuessler:2007av}).}
Using the modified Lagrangian in \eq{V3heff}, the $hh \to hh$ scattering amplitude reads (see also \fig{fig:hscattering})  
\bee
\mathcal{M}=-\lambda_{hhh}^2\left(\frac{1}{s- m_h^2} +\frac{1}{t- m_h^2}+ \frac{1}{u- m_h^2} \right) - \lambda_{hhhh} \, , 
\eee
with $s$, $t$, $u$ denoting the standard Mandelstam variables defined in the center of mass frame. 
In particular, we also have $t = - (s-4m_h^2) \sin^2\frac{\theta}{2}$ and $u = - (s-4m_h^2) \cos^2\frac{\theta}{2}$, 
where $\sqrt{s}$ is the center of mass energy and $\theta$ is the azimuthal angle with respect to the colliding axis.  

The $J=0$ partial wave is found to be
\beq 
\label{a0hhtohh}
a^0_{hh \to hh} = - \frac{1}{2} \frac{\sqrt{s(s-4 m^2_h)}}{16 \pi s} \left[ \lambda_{hhh}^2\left(\frac{1}{s- m_h^2} 
- 2 \frac{\log\frac{s-3m^2_h}{m^2_h}}{s-4m^2_h} \right) + \lambda_{hhhh} \right] \, ,
\eeq
where
we paid attention to keep the kinematical factors which makes the amplitude to vanish at threshold ($\sqrt{s} = 2 m_h$) 
and we multiplied by an extra $1/2$ factor due to the presence of identical particles in the initial and final state 
(see e.g.~\cite{DiLuzio:2016sur} for a collection of relevant formulae). 
Following standard arguments \cite{Lee:1977eg,Luscher:1988gc}, 
perturbative unitarity bounds are obtained by requiring $\abs{\Re{a^0_{hh \to hh}}} < 1/2$.  

\begin{figure}[h!]
\centering
\includegraphics[angle=0,width=7.cm]{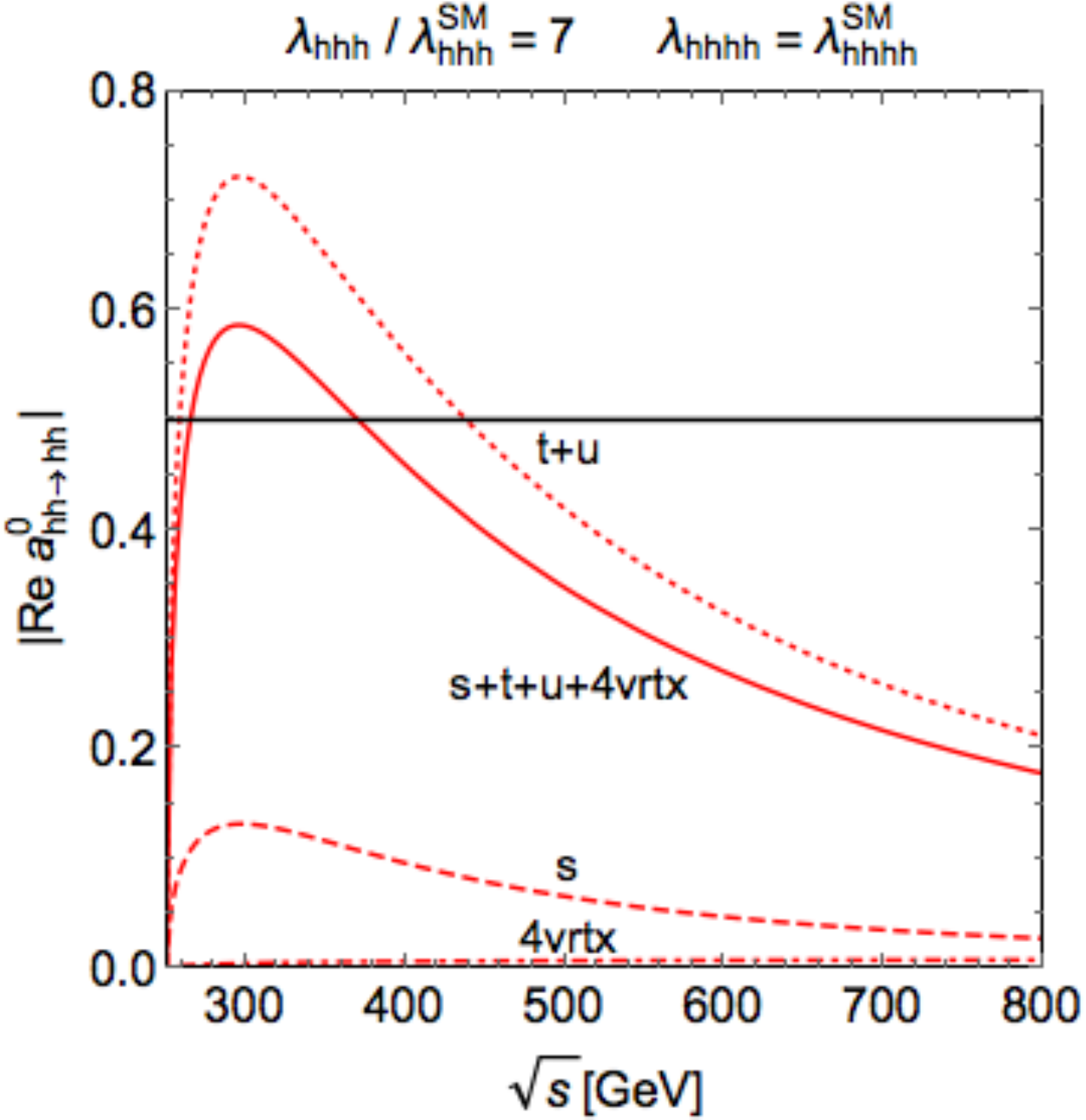} \qquad
\includegraphics[angle=0,width=7.cm]{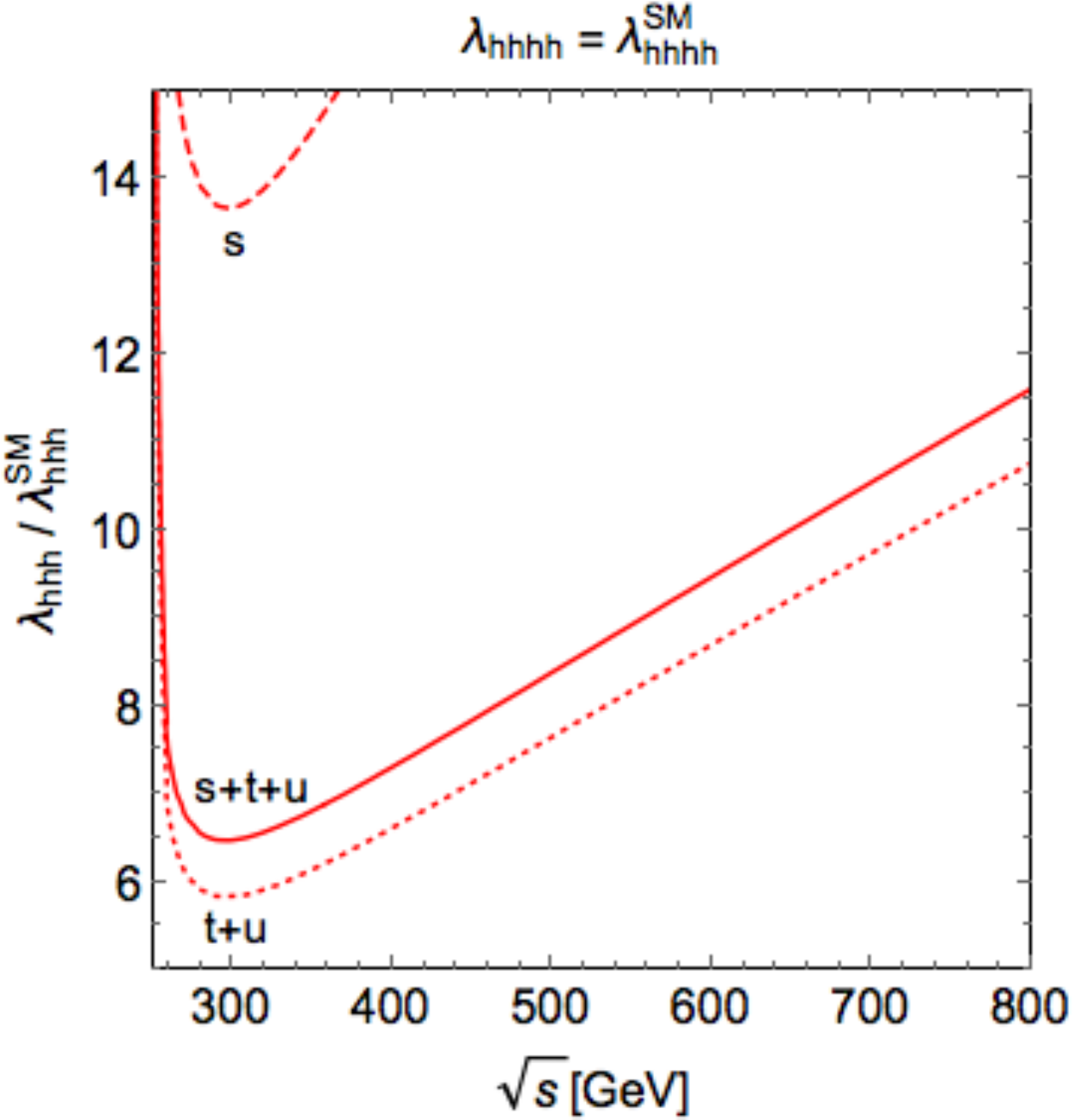} \\
   \vspace*{0.5cm}
\includegraphics[angle=0,width=7.cm]{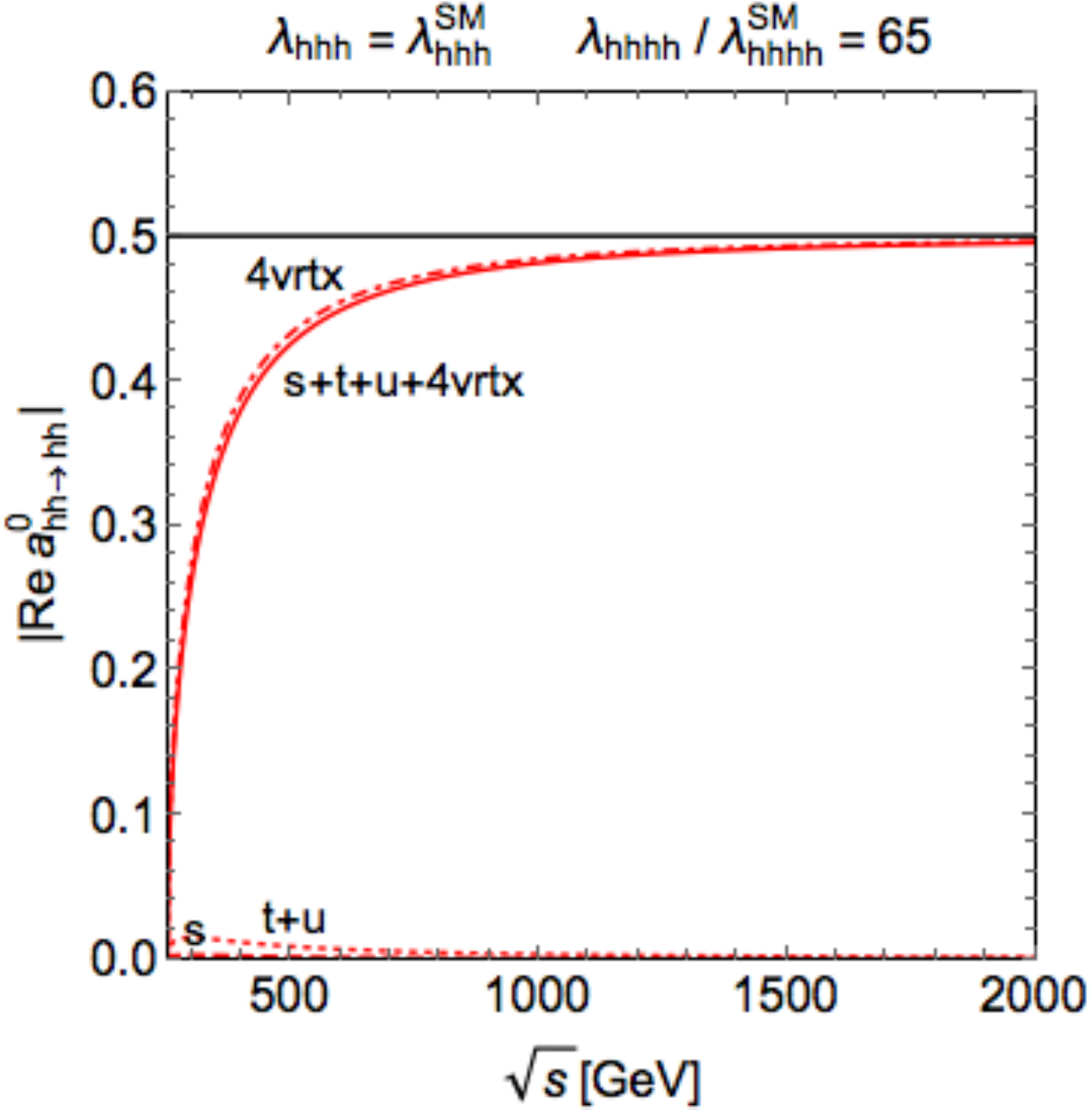} \qquad
\includegraphics[angle=0,width=7.cm]{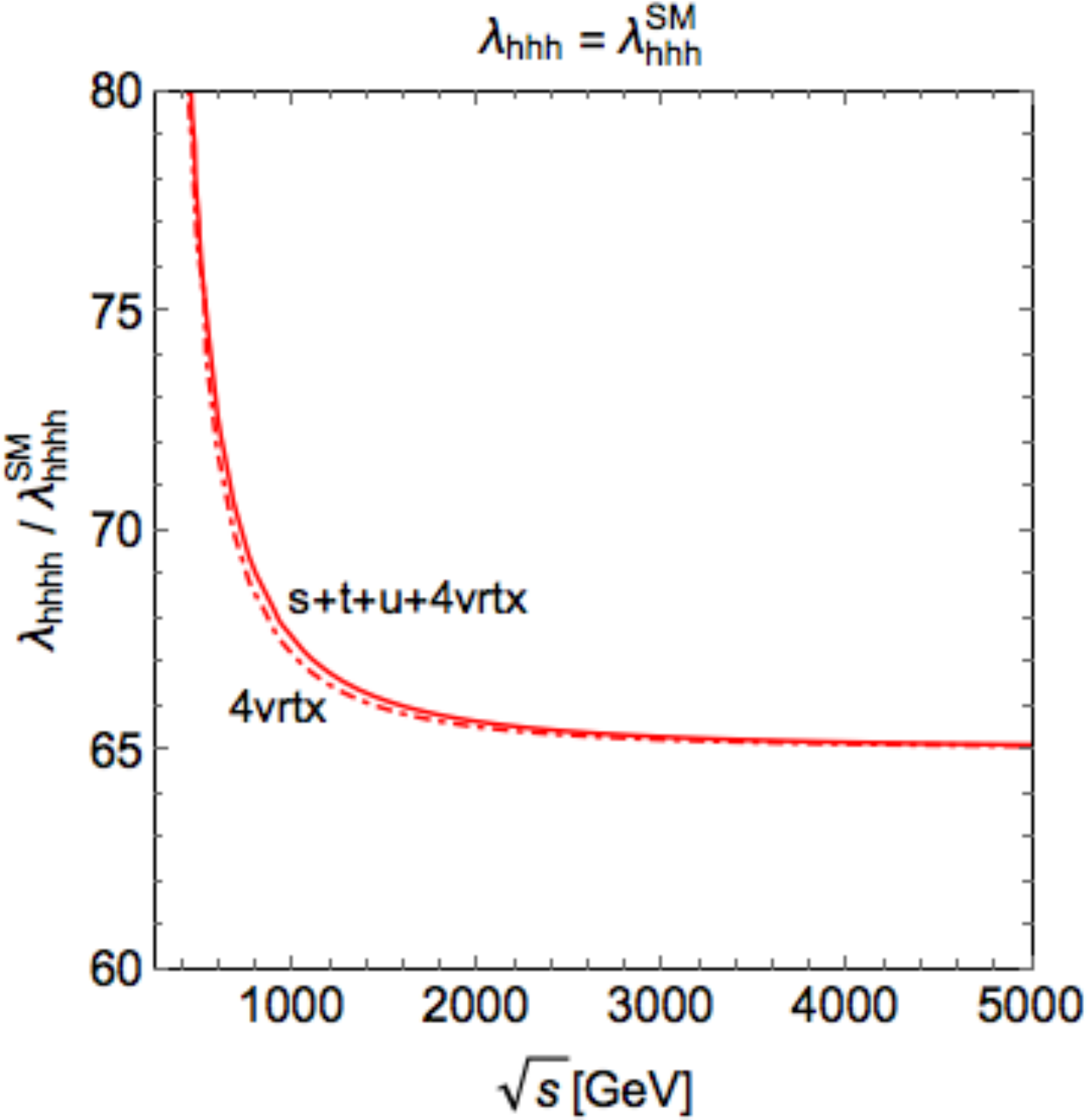}
\caption{\label{fig:a0bound} 
\emph{Up/Left}: Kinematical dependence of $\abs{\Re{a^0_{hh \to hh}}}$ for the reference values 
$\lambda_{hhh} / \lambda_{hhh}^{\rm SM} =7$ and $\lambda_{hhhh} = \lambda_{hhhh}^{\rm SM}$.   
\emph{Up/Right}: Partial-wave unitarity bound $\abs{\Re{a^0_{hh \to hh}}} < 1/2$ on $\lambda_{hhh} / \lambda_{hhh}^{\rm SM}$ 
as a function of $\sqrt{s}$ and for $\lambda_{hhhh} = \lambda_{hhhh}^{\rm SM}$. 
\emph{Down/Left}: Kinematical dependence of $\abs{\Re{a^0_{hh \to hh}}}$ for the reference values 
$\lambda_{hhhh} / \lambda_{hhhh}^{\rm SM} = 65$ and $\lambda_{hhh} = \lambda_{hhh}^{\rm SM}$.   
\emph{Down/Right}: Partial-wave unitarity bound $\abs{\Re{a^0_{hh \to hh}}} < 1/2$ on $\lambda_{hhhh} / \lambda_{hhhh}^{\rm SM}$ 
as a function of $\sqrt{s}$ and for $\lambda_{hhh} = \lambda_{hhh}^{\rm SM}$. 
Dashed, dotted, dot-dashed and full curves denote respectively the $s$, $t+u$, $4$vrtx and $s+t+u+4$vrtx contribution 
to the partial wave. Note that $s$ and $4$vrtx have the opposite sign of $t+u$ (cf.~\eq{a0hhtohh}). 
}
\end{figure}

The bound is displayed in \fig{fig:a0bound} for the orthogonal cases in which either 
$\lambda_{hhh}$ (upper plots) or $\lambda_{hhhh}$ (lower plots) is modified 
with respect to the SM case. Note that the situation is qualitatively different for the two cases: 
being $h^3$ a relevant operator, the unitarity bound on $\lambda_{hhh}$ is maximized 
at low energy, 
while in the case of $h^4$ the partial wave grows with energy 
reaching an asymptotic value at $\sqrt{s} \to \infty$.\footnote{Note that this behaviour is different 
from the case of effective operators, whose scattering amplitudes grow indefinitely 
with the energy.} In particular, from the right-side plots in \fig{fig:a0bound} we read the following 
unitarity bounds
\beq
\label{unitaritybounds}
\abs{\lambda_{hhh} / \lambda^{\rm SM}_{hhh}} \lesssim 6.5 
\qquad \text{and} \qquad
\abs{\lambda_{hhhh} / \lambda^{\rm SM}_{hhhh}} \lesssim 65 \, .
\eeq
Of course, one expects that new physics effects should modify at the same time both 
$\lambda_{hhh}$ and $\lambda_{hhhh}$. However, since the $h^3$ and $h^4$ operators 
dominate the partial wave in two well-separated energy regimes they cannot cancel each other over the 
whole range of $\sqrt{s}$. Hence, since we require perturbativity at any value of $\sqrt{s}$, 
the bounds in \eq{unitaritybounds} hold also in more general situations (as we have checked numerically 
by employing the full expression in \eq{a0hhtohh}).

Let us inspect, for instance, the case where the modified SM potential arises from the operator $\abs{H}^6$ as in \eq{V6}. In such a case we have  
\begin{align}
\lambda_{hhh} &= \lambda^{\rm SM}_{hhh}  + 6 \, c_6 v \simeq \lambda^{\rm SM}_{hhh} \left( 1 + 7.8 \, c_6 \right) \, , \\
\lambda_{hhhh} &= \lambda^{\rm SM}_{hhhh} + 36 \, c_6 \simeq \lambda^{\rm SM}_{hhhh} \left( 1 + 47 \, c_6 \right) \, .
\end{align}
The perturbativity bound coming from the $h^3$ ($h^4$) vertex in \eq{unitaritybounds} translates into 
$\abs{c_6} \lesssim 0.71 \ (1.4)$.

\subsubsection{Loop-corrected vertices}

An alternative way to assess perturbativity is by requiring that the 
loop-corrected trilinear scalar vertex is smaller (in absolute value) than $\lambda_{hhh}$. 
If that were not the case, we clearly could not 
reliably use perturbation theory whenever $\lambda_{hhh}$ entered some physical process.
A similar criterium was employed 
for trilinear scalar interactions in Ref.~\cite{DiLuzio:2016sur}, by setting to zero the external momenta of the 3-point function. 
Following the same argument, we obtain 
\beq 
\label{deltahhhkin}
\Delta \lambda_{hhh} (p_i \to 0) =  \frac{1}{32 \pi^2} \lambda_{hhh}^3 \frac{1}{m_h^2} .
\eeq 
By requiring that $|\Delta \lambda_{hhh}/\lambda_{hhh}|<1$, 
the trilinear Higgs self-coupling is bounded by
\beq 
\abs{\lambda_{hhh} / \lambda_{hhh}^{\text{SM}}} \lesssim 12 \, . 
\label{eq:pert12}
\eeq 
A stronger perturbativity bound can be obtained by looking at the full 
kinematical dependence of the trilinear vertex at the one-loop order. 
Considering the finite one-loop contribution due to $\lambda_{hhh}$ we obtain 
\beq 
\label{deltahhhkin}
\Delta \lambda_{hhh} (\sqrt{s}, m_h) = - \frac{1}{16 \pi^2} \lambda_{hhh}^3 C_0 (m_h^2, m_h^2, s; m_h, m_h, m_h) \, ,
\eeq 
where $C_0$ is a scalar Passarino-Veltman function (defined according to the conventions of Ref.~\cite{Patel:2015tea}) 
and $\sqrt{s}$ denotes the off-shell momentum of a Higgs boson line. 
Since we only took into account the loop correction where the $\lambda_{hhh}$ 
coupling occurs, there are no divergent contributions, 
and we neglected scheme-dependent finite terms.
It should be understood that what we aim at is not a proper calculation of the quantum corrections 
to $\lambda_{hhh}$, but rather a simple estimate of the validity of perturbation theory. 
The reason why an estimate based solely on the 
contribution in \eq{deltahhhkin} is reasonable is the following: 
$i)$ in the large $\lambda_{hhh}$ limit, where the perturbativity bound is relevant, 
pure SM contributions are subleading 
and $ii)$ even though by gauge invariance one should worry about 
simultaneous $\lambda_{hhhh}$ corrections,  
these are divergent and hence scheme dependent. Then, the estimate 
in Eq.~(34) would be inaccurate only if the finite contribution (in a given renormalization scheme) due to $\lambda_{hhhh}$ 
were to cancel the one stemming from $\lambda_{hhh}$ to a large extent and over the full kinematical range. 
This however is very unlikely, given that the corrections have a very different kinematical dependence. 

\begin{figure}[h!]
\centering
\includegraphics[angle=0,width=7.cm]{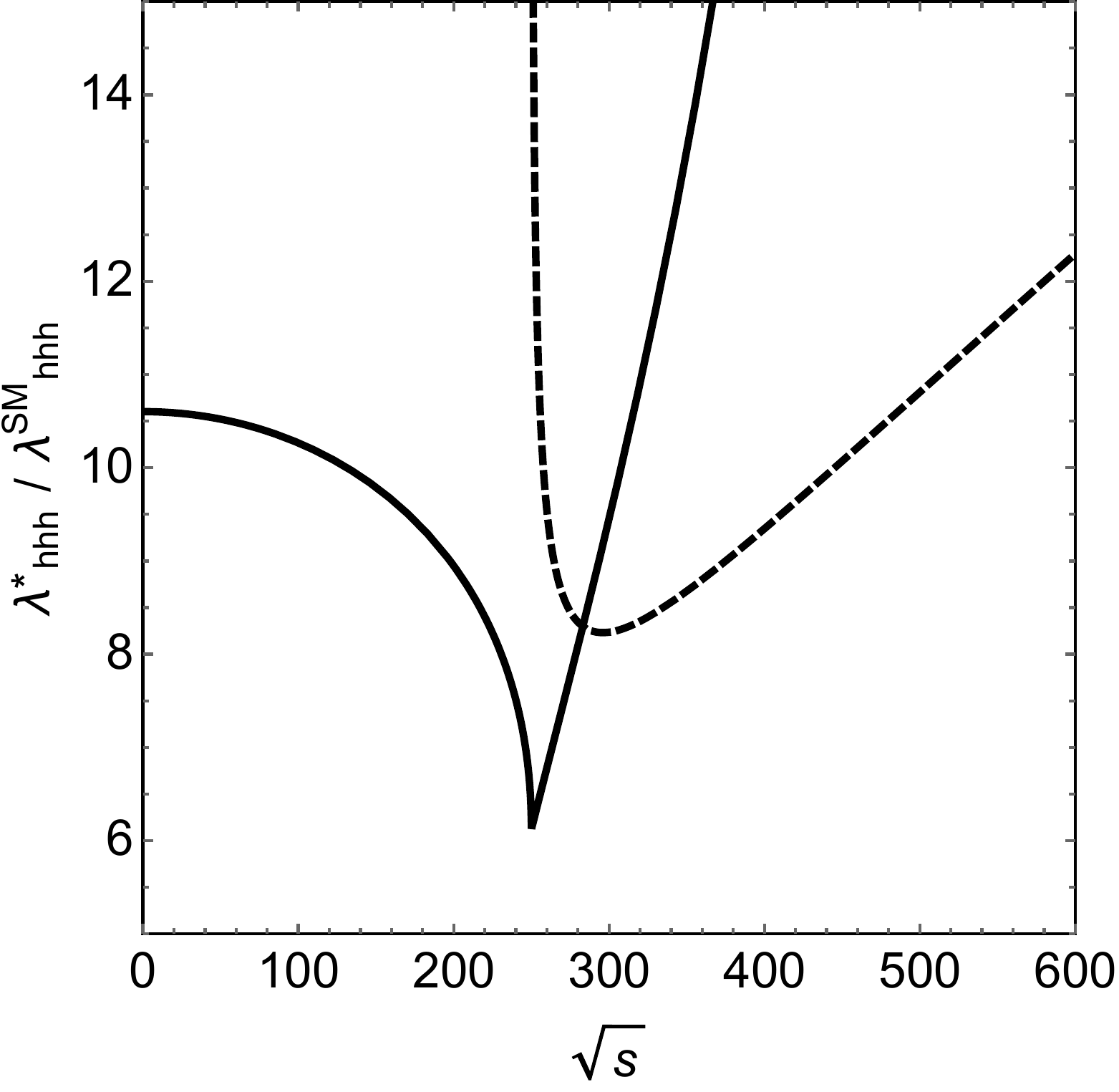}
\caption{\label{fig:Pertboundhhh} 
Perturbativity bound $\lambda_{hhh} < \lambda^*_{hhh}$ from the loop-corrected 
trilinear vertex as a function of $\sqrt{s}$. Full and dashed curves denote respectively the real 
($\abs{\Re (\Delta\lambda_{hhh}) / \lambda_{hhh}} < 1$) and imaginary ($\abs{\Im (\Delta\lambda_{hhh}) / \lambda_{hhh}} < 1$) contributions to the bound due to the vertex correction in \eq{deltahhhkin}. 
}
\end{figure}

The perturbativity bound, denoted by $\lambda^*_{hhh}$, is shown in \fig{fig:Pertboundhhh} as a function of $\sqrt{s}$. 
Note that above threshold, $\sqrt{s} > 2 m_h$, 
$C_0$ develops an imaginary part and hence we have separately considered both the real and imaginary contribution to the bound. 
Since one should require that perturbativity must hold for any value of $\sqrt{s}$, the bound is maximized close to threshold and reads 
\beq 
\label{pertboundhhhmax}
\abs{\lambda_{hhh} / \lambda_{hhh}^{\text{SM}}} \lesssim 6 \, , 
\eeq
which is consistent with the (conceptually different) constraint obtained in \eq{unitaritybounds}. 

A similar argument can be used to set a perturbativity bound on $\lambda_{hhhh}$ by looking at its beta function 
(see e.g.~\cite{Goertz:2015nkp}). 
By requiring $\abs{\beta_{\lambda_{hhhh}}/\lambda_{hhhh}}<1$, we get $\abs{\lambda_{hhhh}} < \frac{16 \pi^2}{3} \simeq 53$. 
Normalizing the latter with respect to the SM value implies 
\beq 
\label{lamhhhhbf}
\abs{\lambda_{hhhh} / \lambda^{\rm SM}_{hhhh}} \lesssim 68 \, ,
\eeq
which again is consistent with \eq{unitaritybounds}.

In the end, given the impossibility of setting genuine model-independent bounds on $\lambda_{hhh}$ beyond perturbativity, 
we focus in the next section on UV complete scenarios when investigating the question of the maximal value of the triple Higgs coupling. 
We focus for simplicity on weakly coupled models, as they retain a higher degree of predictivity and we have full control of the theory.

\section{UV complete models}
\label{UVmodels}

\begin{figure}[h!]
\centering
\vspace{1cm}
\includegraphics[angle=0,width=5.5cm]{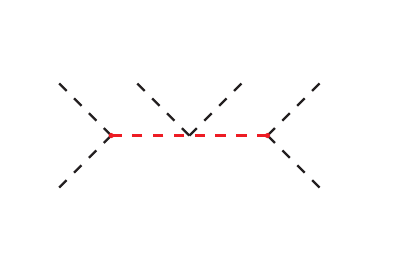} 
\hspace*{-1cm}
\includegraphics[angle=0,width=5.5cm]{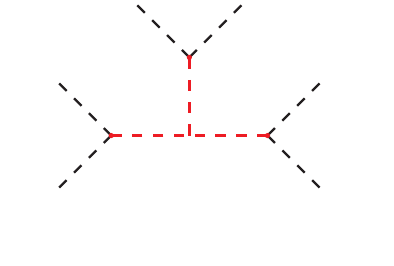}  
\hspace*{-1cm}
\includegraphics[angle=0,width=5.5cm]{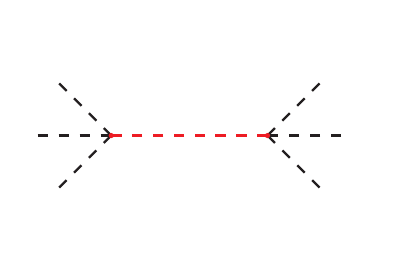}  
\caption{\label{fig:dia} 
Tree-level generation of the $\abs{H}^6$ operator (external lines, black) 
obtained by integrating out new scalar degrees of freedom (internal propagators, red).}
\end{figure}

If the new degrees of freedom are very light, they can affect the Higgs-pair production process in different ways 
(like e.g.~resonant production \cite{Dolan:2012ac,No:2013wsa,Chen:2014ask,Martin-Lozano:2015dja,Godunov:2015nea,Banerjee:2016nzb,Huang:2017jws,Lewis:2017dme}
or by scalar/fermionic contributions to the gluon fusion loop~\cite{Dawson:2015oha, Grober:2016wmf, Agostini:2016vze}) and the dominant effect does not need to be associated with the $\lambda_{hhh}$ 
coupling deviation. Hence, we focus on the case where the new physics is above the EW scale, 
but not necessarily yet in the EFT regime where the effects are expected to decouple rapidly. 
The latter language is nonetheless useful in order to classify the representations 
which are potentially more prone to induce a large effect: at tree level there are basically three class of diagrams 
(cf.~\fig{fig:dia}) which can generate $\abs{H}^6$ by integrating out a heavy new scalar degree of 
freedom.\footnote{Note that it is also possible to exchange a massive vector at tree level, e.g.~in presence of the 
trilinear coupling $g_V H^\dag D_\mu H \, V^\mu$, where $V^\mu$ has gauge quantum numbers $(1,1,0)$ or $(1,3,0)$ 
(see e.g.~\cite{delAguila:2010mx,Biggio:2016wyy}). After integrating $V^\mu$ out and applying the equations of motion 
one obtains an $\abs{H}^6$ operator with Wilson coefficient proportional to $\lambda g_V^2 / M^2_V$. 
On the other hand, massive vectors (either in their gauge extended of strongly coupled version) require a UV completion, 
thus going beyond our simplifying assumption of a one-particle extension of the SM. 
} 
Here, we concentrate on trilinear Higgs self-coupling modifications generated by $\abs{H}^6$, since they uniquely
modify the Higgs self-couplings. Also the operator $\partial_{\mu}(H^{\dagger}H)\partial^{\mu}(H^{\dagger}H)$ 
gives a contribution to the shift in the trilinear Higgs self-coupling, 
but it modifies all other Higgs couplings as well.

In fact, the connecting motive between the diagrams in Fig.~\ref{fig:dia} turns out to be a tadpole operator of the type $\mathcal{O}_\Phi = \Phi f(H)$, 
where $f(H)$ is a string of Higgs fields (or their charged conjugates). The full list of scalar extensions that couple linearly to $H$ can be found 
in \Table{newscalarsHHH} (see also Refs.~\cite{deBlas:2014mba,DiLuzio:2015oha,Jiang:2016czg}), where hyper-chargeless multiplets are understood to be real. 
For simplicity, we will focus on one-particle extensions of the SM in order to point out their features in a clear way.  
\begin{table}[h!]
\renewcommand{\arraystretch}{1.4}
\centering
\begin{tabular}{@{} |c|c|
@{}}
\hline
$\Phi$ &  $\mathcal{O}_\Phi$ 
\\ 
\hline
$(1,1,0)$ & $\Phi H H^\dag$ 
\\
$(1,2,\tfrac{1}{2})$ & $\Phi H H^\dag H^\dag$ 
\\
$(1,3,0)$ & $\Phi H H^\dag$ 
\\
$(1,3,1)$ & $\Phi H^\dag H^\dag$ 
\\
$(1,4,\tfrac{1}{2})$ & $\Phi H H^\dag H^\dag$ 
\\
$(1,4,\tfrac{3}{2})$ & $\Phi H^\dag H^\dag H^\dag$ 
\\
\hline
  \end{tabular}
  \caption{\label{newscalarsHHH} 
  List of new scalars $\Phi$ 
  inducing a tree-level modification of the triple-Higgs coupling    
  via the tadpole operator $\mathcal{O}_\Phi$. 
}
\end{table}

Another useful way to understand the origin of the trilinear Higgs self-coupling modification, which does not rely on the EFT language is the 
following: the tadpole operator will unavoidably generate a vev for $\Phi$, and the 
neutral components $h^0 \subset H$ and $\phi^0 \subset \Phi$ will mix via the tadpole operator itself. 
After projecting the two neutral components on the Higgs boson mass eigenstate, namely $h^0 \to h \cos\theta$ and $\phi^0 \to h \sin\theta$,
we have the following contribution to the triple-Higgs vertex
\beq 
\Delta \lambda_{hhh} = \mu_{\Phi} \sin\theta \cos^2\theta 
\qquad \text{or} \qquad
\lambda_{\Phi} v \sin\theta \cos^3\theta \, ,
\eeq 
depending whether the tadpole operator is $d=3$ ($\mu_{\Phi}$ coupling) or $d=4$ ($\lambda_{\Phi}$ coupling). 
Since there is a single suppression from the mixing angle, bounded at the level of $\theta \lesssim 0.3$ from Higgs coupling measurements, 
the tadpole interaction is expected to yield the largest contribution, while other mixing operators in the scalar potential entail extra suppressions 
from $\sin\theta$.
We can also naively estimate the contribution in the following way: 
assuming that $\mu_{\Phi}/v \lesssim 4 \pi$ and $\lambda_\Phi \lesssim 4 \pi$ by perturbativity we get 
\beq 
\frac{\Delta \lambda_{hhh}}{\lambda^{\text{SM}}_{hhh}} \lesssim 4 \pi \sin\theta \cos^2\theta \frac{v^2}{3 m_h^2} \sim 4 \, . 
\eeq 
To make this estimate more precise, we will look in detail at two paradigmatic examples among those in \Table{newscalarsHHH}:  
one model which exhibits a tree-level custodial symmetry (singlet case, \sect{sec:singlet}) and one which does not (triplet case, \sect{sec:triplet}). 

A notable feature of tadpole interactions is that, being ``odd'' in $\Phi$, they are potentially bounded by vacuum stability considerations. 
Remarkably, we find that vacuum stability is never a crucial discriminant for bounding the largest value of $\lambda_{hhh}$, because whenever 
the tadpole coupling is large the instability can be tamed by large (within the perturbativity domain) quartic couplings. 
For this reason we find it relevant to discuss in \sect{looptrilinear} a class of loop-induced trilinear Higgs self-couplings 
that arise due to vector-like fermions, where one can establish a direct connection between $\lambda_{hhh}$ and the vacuum instability.

\subsection{Tree-level custodially symmetric cases}
\label{sec:singlet}

Among the cases in \Table{newscalarsHHH}, the singlet and the doublet do not violate custodial symmetry at tree level and hence 
have the chance to yield the largest contribution to $\lambda_{hhh}$. We will discuss in detail the singlet case, while we only comment on the case of the 
doublet towards the end of the subsection. 
The scalar potential reads 
\begin{equation}
V(H,\Phi) = 
\mu_1^2 |H|^2+\lambdah |H|^4+\frac{1}{2} \mu_2^2 \Phi^2+\mu_4 |H|^2 \Phi +\frac{1}{2} \lambda_3 |H|^2 \Phi^2+\frac{1}{3} \mu_3 \Phi^3+\frac{1}{4} \lambda_2 \Phi^4 \, ,\label{eq:Lsinglet}
\end{equation}
where we have omitted a tadpole term for the singlet field, as it can be reabsorbed in the singlet vev by a field redefinition.  

In fact, the $\mu_4$ coupling unavoidably induces a vev for $\Phi$ and also leads to a mixing between $H$ and $\Phi$. 
In \app{app:singlet} we give the tadpole equations and we define the mixing angle $\theta$ between the singlet and doublet fields. Some of the parameters of the potential can be expressed in terms of the physical masses and vevs and their mixing angle. We chose as input parameters
\beeq
\label{inputparsinglet}
\vh=246.2 \text{ GeV}\,, \; \; \; \vs\,, \; \; \; m_1=125 \text{ GeV}\,, \; \; \; m_2\,, \; \; \; \theta\,, \; \; \;\lambda_{2}\,, \; \; \; \lambda_3\,.
\eeeq
Their relations to the other parameters of the potential can be found in \app{app:singlet}. 
Note that the scenario in which the SM-like Higgs boson is heavier than the singlet-like scalar is phenomenologically viable as well, 
but we will restrict ourselves to the case $m_1 \ll m_2$. The reason being that we want to discuss deviations to the Higgs pair production process that are mainly stemming from the trilinear Higgs self-coupling, 
while the contribution from the exchange of the singlet-like Higgs boson in the triangle diagrams is suppressed. For discussion on resonant Higgs pair production in the singlet model we refer to Refs.~\cite{Dolan:2012ac,No:2013wsa,Chen:2014ask,Martin-Lozano:2015dja,Godunov:2015nea,Banerjee:2016nzb,Huang:2017jws,Lewis:2017dme}. 

The trilinear Higgs self-coupling is given by
\be
\label{lamhhhsinglet}
\lambda_{hhh}&=&6 \lambdah \vh \cos^3\theta - (3 \mu_4 + 3 \lambda_3 \vs) \cos^2\theta \sin\theta 
    + 3 \lambda_3 \vh \cos\theta \sin^2\theta - \sin^3\theta (2 \mu_3+6 \vs \lambda_2) \nonumber \\&=& \lambda_{hhh}^{\rm SM} \cos\theta \left[ 1 + \sin^2\theta \left( \frac{\lambda_3 \vh^2}{m_1^2}-1\right)+ \sin^4\theta \frac{\vh^2}{3 \vs^2}\left(1-\frac{m_2^2}{m_1^2}\right)\right. \nonumber \\ 
    &-&\left.\frac{\vh}{3 \vs}\frac{\sin^3\theta}{\cos\theta} \left(2 \sin^2\theta +2 \cos^2\theta \frac{m_2^2}{m_1^2}-\frac{\lambda_3 \vh^2}{m_1^2}+\frac{2 \vs^2\lambda_2}{m_1^2}\right)\right] \,,  
\ee
where in the last step we expressed $\lambda_{hhh}$ in terms of the input parameters in \eq{inputparsinglet}. 

In order to make contact with the discussion at the beginning of \sect{UVmodels} on the importance of tadpole operators 
for enhancing the trilinear Higgs self-coupling, let us compare the expression in \eq{lamhhhsinglet} with the 
one obtained in the $\mathbb{Z}_2$-symmetric limit with $\mu_{3,4} \to 0$, which yields
\bee
\lambda_{hhh}^{\mathbb{Z}_2\text{--symmetric}}=\lambda_{hhh}^{\rm SM} \left(\cos^3\theta -\sin^3 \theta \frac{\vh}{\vs} \right)\,.
\eee
It is thus evident that the shift in the trilinear Higgs self-coupling can be much larger for the 
general singlet potential with tadpole 
terms. In the last step of \eq{lamhhhsinglet} 
we see indeed that potentially large contributions can arise from sizable values of $\lambda_3$.\footnote{For comparison, in the $\mathbb{Z}_2$-symmetric case one finds that the maximal deviations 
on the trilinear Higgs self-coupling are at the $10\%$ level, in the case where 
the second Higgs boson cannot be directly detected at the LHC \cite{Gupta:2013zza, Efrati:2014uta}.}

In the following we will discuss which values the trilinear Higgs self-coupling can take, by accounting for several constraints.

\subsubsection{Indirect bounds}
\label{sec:indirectbounds}
The model parameters can be restricted by EW precision tests, Higgs coupling measurements, perturbativity arguments and vacuum stability. 
These will then indirectly constrain the trilinear Higgs self-coupling in the model.
\\
\\
\underline{EW precision tests:}
\\
In Ref.~\cite{Lopez-Val:2014jva} it was pointed out that the measurement of the $W$ boson mass constrains the scalar singlet model more strongly 
than a fit on the $S$, $T$, $U$ parameters. Even though the study in Ref.~\cite{Lopez-Val:2014jva} concerns a $\mathbb{Z}_2$ symmetric potential, 
we can use the bounds here, since at the one-loop order the additional parameters in the scalar potential 
do not play any role for the gauge boson vacuum polarizations. For $m_2>800\text{ GeV}$, Ref.~\cite{Lopez-Val:2014jva} finds 
the bound $|\sin\theta|<0.2$. 
\\
\\
\underline{Higgs coupling measurements:}
\\
The Higgs production and decay rates are modified with respect to the SM by a universal factor
\begin{align}
\sigma(pp\to h+X)&=\cos^2\theta\,\sigma_{\rm SM}(pp\to h+X) \, , \\ 
\Gamma(h\to XX)&=\cos^2 \theta \,\Gamma_{\rm SM}(h\to XX) \, .
\end{align}
If the SM-like Higgs boson corresponds to the lightest eigenstate, its branching ratios 
are not modified compared to the SM.
In Ref.~\cite{Aad:2015pla} a limit on $\sin^2\theta<0.12$ at 90\% C.L. from Higgs signal measurements is given. 
This limit turns out to be stronger than the limits from direct searches of the heavier Higgs boson, 
as long as $m_2 > 450 \ \text{GeV}$ \cite{Robens:2016xkb}, such that we will not need to take 
the latter into account for the parameter space we consider. 
\\
\\
\underline{Perturbativity:}
\\
For large enough potential couplings unitarity is violated in tree-level scattering processes, 
thus signalling the breakdown of perturbation theory.
Simple criteria can be derived from the $ii \to jj$ scattering, with $i$ and $j$ running over the (real) Higgs and singlet fields. 
By requiring $\abs{\text{Re} \, a_0} < 1/2$ for the eigenvalues of the $J=0$ partial-wave scattering matrix, we derive the following constraint 
in the high-energy limit 
\bee
3 \left(\lambdah+\lambda_2\right) \pm \sqrt{9\left(\lambdah-\lambda_{2}\right)^2+\lambda_3^2} < 16 \pi \,.
\label{eq:unitbound}
\eee
The dimensionful parameters $\mu_3$ and $\mu_4$ can be restricted by unitarity arguments as well. 
However, being associated to super-renormalizable operators the bounds are maximized at low energies,  
where the possible presence of resonances actually requires a careful treatment of the pole singularities. 
Following the argument of Ref.~\cite{DiLuzio:2016sur}, in order to define the perturbative domain of $\mu_3$ and $\mu_4$ we require 
instead that the one-loop corrected trilinear scalar couplings at zero external momenta 
remain smaller than the tree-level ones. In the $SU(2)$ limit we obtain
\bee
\frac{\abs{\mu_4}}{\text{max}\, \left(\abs{\mu_1}, \abs{\mu_2}\right)} < 4 \pi \, ,
\quad\land\quad 
\abs{\frac{\mu_3}{\mu_2}} < 4 \pi \,. \label{eq:boundAmu}
\eee 
The saturation of the bounds in \eqs{eq:unitbound}{eq:boundAmu} correspond to an extreme situation, 
where we progressively enter a strongly-coupled regime for which the perturbative calculation does 
not make sense anymore. 
For this reason, we will also present the results in another regime where we keep the couplings significantly smaller.  For that we use in \eq{eq:boundAmu} the replacement $4 \pi \to 1$ and in the scan we restrict $0<\lambda_2 <1/6$ and $|\lambda_3|<1$.
\\
\\
\underline{Vacuum stability:}
\\
The requirement that the scalar potential is bounded from below imposes the following conditions on the quartic scalar interactions
\bee
\lambdah>0 \, , \quad\land\quad 
\lambda_{2}>0 \, , \quad\land\quad
\lambda_3>-2 \sqrt{\lambda_{2}\lambdah} \,. \label{eq:boundedness}
\eee
The study of the minima of the scalar potential exhibits a rich structure, with new local minima 
(e.g.~in $h=0$) that arise in some regions of the parameter space and which might 
eventually destabilize the EW vacuum. A detailed analysis of the vacuum structure at tree level can be found in 
Refs.~\cite{Espinosa:2011ax,Chen:2014ask}. 
We check for vacuum stability by using {\tt Vevacious} \cite{Camargo-Molina:2013qva, Camargo-Molina:2014pwa}, 
with a model file generated with {\tt SARAH} \cite{Staub:2008uz, Staub:2009bi, Staub:2010jh, Staub:2012pb, Staub:2013tta}.

\subsubsection{Results}
In order to show the results we perform a scan over the parameter space. The universally scanned parameters 
in both the cases are
\bee
m_1=125\text{ GeV}, \hspace*{0.5cm}\;800\text{ GeV}< m_2 < 2000 \text{ GeV}, \; \\ \vh=246.2\text{ GeV},  \hspace*{0.5cm}\;  |\vs|<m_2, \hspace*{0.5cm}\; 0.9 < \cos\theta < 1\,. \nonumber
\eee 
We will perform two different scans.
In the first one we use the maximally allowed values according to the perturbativity argument
\bee
\begin{split}
\text{Scan 1: } \hspace*{0.5cm} 0<\lambda_2<\frac{8}{3} \pi, \hspace*{0.5cm} \; |\lambda_3|<16\pi,
\end{split}
\eee
and reject all points that do not fulfil \eq{eq:unitbound}, \eq{eq:boundAmu} and \eq{eq:boundedness}.
In the second scan we restrict ourselves to a weakly-coupled scenario and scan the input parameters 
\bee
\begin{split}
\text{Scan 2: } \hspace*{0.5cm} 0<\lambda_{2}<1/6,\hspace*{0.5cm} \; |\lambda_3|<1,
\end{split}
\eee
together with $|\mu_4|/\text{max}(|\mu_1|, |\mu_2|)<1$ and $|\mu_3/\mu_2 |<1$. 
\begin{figure}
\hspace*{-0.4cm}\includegraphics[width=8.5cm]{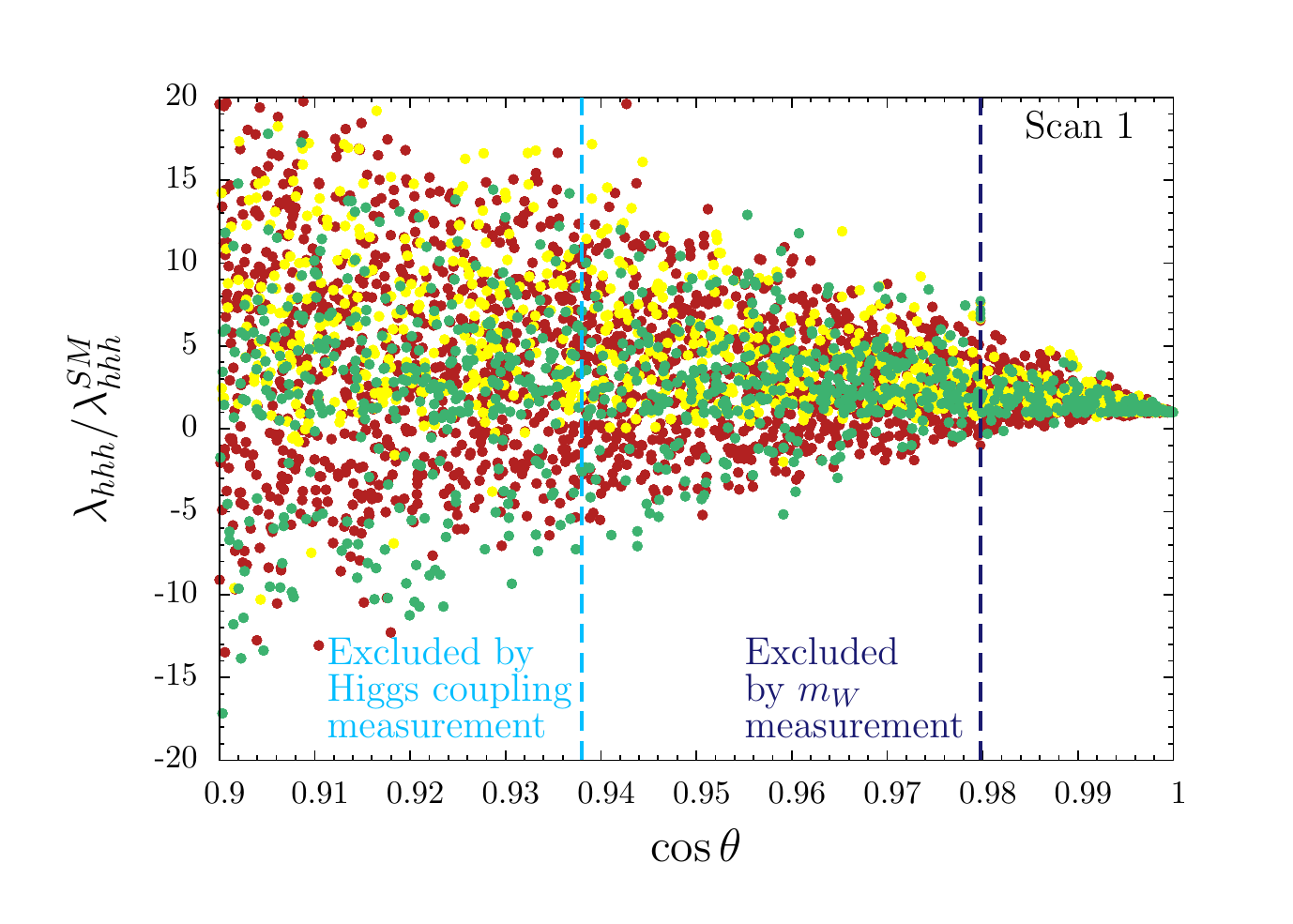}\hspace*{-0.4cm}
\includegraphics[width=8.5cm]{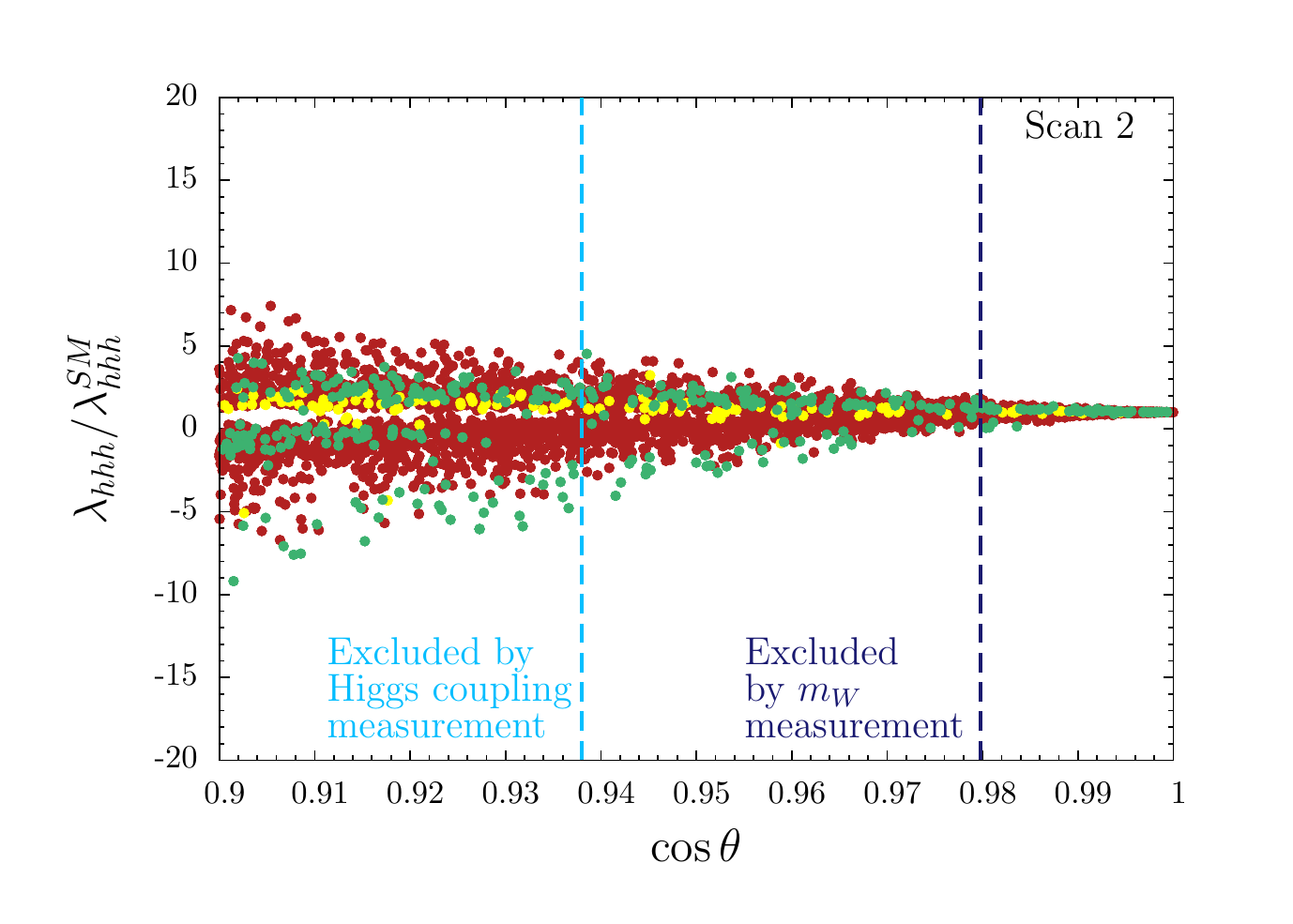}
\caption{{\it Left:} The trilinear Higgs self coupling normalised to the SM reference value for scan 1 (strongly-coupled regime). 
The red/yellow/green points correspond respectively to unstable/metastable/stable configurations. 
The dashed vertical lines indicate the bounds on $\cos \theta$ from the respective experimental measurements. 
{\it Right:} Same as in left panel, but for scan 2 (weakly-coupled regime). 
}\label{fig:singletresult}
\end{figure}

In \fig{fig:singletresult} the trilinear Higgs self coupling normalised to the SM coupling is shown. The color code of the points indicate whether they correspond to a stable, metastable or unstable vacuum configuration. By accounting for the bounds of the $m_W$ boson measurement we find the following range for the allowed trilinear Higgs self-coupling: 
\bee
&\text{Scan 1:} \hspace*{0.5cm} & -1.5 < \lambda_{hhh}/\lambda_{hhh}^{\rm SM} < 8.7 \, , \\
&\text{Scan 2:} \hspace*{0.5cm} & \,-0.3 < \lambda_{hhh}/\lambda_{hhh}^{\rm SM} < 2.0 \, .
\eee 
In fact, the largest value of the trilinear Higgs self-coupling is crucially related to the perturbativity domain. The bounds on the trilinear Higgs self-coupling obtained from scan 1 should hence be treated with care, as they are very close to the non-perturbative regime and loop corrections can be expected to be large.
This can be easily understood looking at the formulae in \eq{lamhhhsinglet}. By allowing for rather large values of e.g.~$\lambda_3$ we can get much larger deviations. Note that we find here a larger value for $ \lambda_{hhh}/\lambda_{hhh}^{\rm SM}$ as in \sect{sectperthhh}, since we require a weaker perturbativity criterium in \eq{eq:boundAmu}, corresponding to the one in Eq.~\eqref{eq:pert12}. Indeed, due to the possible presence of resonances  which requires a careful treatment of the pole singularities we could not apply the bound in \eq{unitaritybounds} from partial-wave unitarity in a straightforward manner. 
On the other hand, as it can be inferred from \fig{fig:singletresult}, the requirement of a stable vacuum has only a very small impact on the bound of the trilinear Higgs self-coupling. 
The little impact of vacuum stability can be understood by the fact that the presence of many parameters in 
the scalar potential basically uncorrelates the stability conditions from the value of the trilinear Higgs self-coupling.

At this point, we would like to comment on previous studies in the context of the scalar singlet. 
In Ref.~\cite{Buttazzo:2015bka}, deviations for $\lambda_{hhh}/\lambda_{hhh}^{\rm SM}$ up to $-10$ were found. Note however that much weaker limits on the mixing angle $\theta$ were employed, since the bound stemming from the $m_W$ measurement was not used. In addition, weaker bounds from the Higgs coupling measurements were employed. 
In Ref.~\cite{Kanemura:2016lkz, Camargo-Molina:2016moz} one-loop corrections to the trilinear Higgs self-coupling were computed. 
They can give large corrections (even up to 100\%) from non-decoupling effects in the Higgs boson loops if $\lambda_3 \vh^2 \gg \mu_2^2$ \cite{Kanemura:2004mg}. 
This is not surprising, given the fact that one is saturating the perturbativity limit where loop effects are not under control. 

We conclude with a few remarks on the other custodial symmetric case, namely the two-Higgs doublet model (2HDM). 
The question of the trilinear Higgs self-coupling was addressed in detail in the context of the $\mathbb{Z}_2$ 
symmetric case \cite{Bernon:2015wef,Baglio:2014nea}, where it was shown that 
the expected deviations are well below those allowed in the general singlet model.
On the other hand, a full study in the context of the general 2HDM (including the $\Phi H H^{\dagger} H^{\dagger}$ 
tadpole operator) is still missing to our knowledge (see however \cite{Ginzburg:2015yva} for a qualitative study). 
In such a case we expect potentially large deviations. 
We leave this study for future investigations.  

\subsection{Tree-level custodially violating cases}
\label{sec:triplet}

We shall discuss the cases corresponding to the last four rows in \Table{newscalarsHHH} altogether, 
since they have in common the fact that the tadpole term 
$\Phi f(H)$ contributing to a potentially sizable triple Higgs self-coupling generates a custodial-breaking 
vev for $\Phi$, which is strongly bounded by EW precision tests. 

Let us exemplify the analysis for the case of a real EW triplet 
with zero hypercharge, $\Phi \sim (1,3,0)$. The scalar potential reads (see e.g.~\cite{Chen:2006pb}) 

\beq 
\label{sptriplet}
V(H,\Phi) = \mu_1^2 \abs{H}^2 + \frac{1}{2} \mu_2^2 \abs{\Phi}^2 + \lambda_1 \abs{H}^4 
+ \frac{1}{4} \lambda_2 \abs{\Phi}^4 + \frac{1}{2} \lambda_3 \abs{H}^2 \abs{\Phi}^2 + \mu_4 H^\dag \sigma^\alpha H \Phi^\alpha \, ,
\eeq
where, without loss of generality, we can take $\mu_4>0$ by reabsorbing the sign in the definition of $\Phi$.
The minimization of the potential and the calculation of the scalar spectrum is deferred to \app{app:triplet}. 
In particular, we can choose the following independent observables as parameter inputs for the model 
\beeq
\label{scanpartripletmain}
\vh = \sqrt{v^2 - 4\vt^2} \,, \; \; \; \vt < 3.5 \ \text{GeV} \,, \; \; \; m_1=125 \text{ GeV}\,, \; \; \; m_2\,, \; \; \; m_{h^\pm}\,, \; \; \;\theta \,, 
\eeeq
where $v = 246.2$ GeV. 
The trilinear Higgs self-coupling is given by
\be
\label{lamhhhtriplet}
\lambda_{hhh}&=& 
6 \lambda_1 \vh \cos^3\theta + 3 \left( \mu_4 - \lambda_3 \vt \right) \cos^2\theta \sin\theta +3 \lambda_3 \vh \cos\theta \sin^2\theta - 6\lambda_2 \vt \sin^3\theta
 \\ 
   &=& \frac{3 m_1^2}{ \vh} \cos\theta 
 \left[ 1 + \left( \frac{2 m_{h^\pm}^2 \vh^2}{(\vh^2+ 4 \vt^2)m_1^2} - 1 \right) \sin^2\theta 
+ \left( \frac{m_{h^\pm}^2 \vh^2 }{(\vh^2+ 4 \vt^2) m_1^2 }  - 1 \right)\frac{\vh}{\vt}  \frac{\sin^3\theta}{\cos\theta}  
 \right] \, , \nonumber
\ee
where in the last step we expressed $\lambda_{hhh}$ in terms of the parameters in \eq{scanpartripletmain}.

\subsubsection{Indirect bounds}
\label{sec:indbounds3}

As in the singlet case, we are going to consider in turn EW precision tests, Higgs coupling measurements,
perturbativity arguments and vacuum stability in order to constrain the trilinear Higgs self-coupling in the triplet model. \\
\\
\underline{EW precision tests:}
\\
The main bound comes from the tree-level modification of the $\rho$ parameter. 
In the SM the custodial symmetry of the Higgs potential ensures the tree-level relation 
$\rho \equiv m_W^2 / m_Z^2 \cos^2\theta_W =1$. 
Extra sources of custodial symmetry breaking which cannot be accounted within the SM
are described by the $\rho_0 \equiv \rho / \rho_{\rm{SM}}$ parameter. 
Provided that the new physics which yields $\rho_0 \neq 1$ does not significantly affect 
the SM radiative corrections,\footnote{This does not need to be the case in models with $\rho \neq 1$ at tree level, 
where four input parameters (instead of three) are required for the 
EW renormalization \cite{Lynn:1990zk,Blank:1997qa,Chen:2006pb}. 
An investigation of this issue is however beyond the scope of this paper.
} 
a global fit to EW observables yields $\rho_0^{\rm (fit)} = 1.00037 \pm 0.00023$ \cite{Olive:2016xmw}. 
In the triplet model one has  
\beq 
\rho^{\rm{tree}}_0 = 1 + 4 \frac{\vt^2}{\vh^2} \, , 
\eeq
and using the $2\sigma$-level bound from $\rho_0^{\rm (fit)}$ we obtain $\vt < 3.5$ GeV. \\
\\
\underline{Higgs coupling measurements:}
\\
In case of a triplet, the Higgs couplings are modified by $\cos\theta$, while the gauge-Higgs boson couplings get a contribution from the triplet admixture 
proportional to $\sin\theta$. The mixing angle between the doublet and triplet scalar fields is necessarily rather small since 
$\theta \to 0$ for $\vt/\vh \to 0$. 
This means that the tree-level Higgs couplings to fermions and gauge bosons are basically unmodified. The charged Higgs boson contributes to the loop-induced $h\to \gamma\gamma$  and $h\to Z \gamma$ decay. Its contribution is however negligible for 
$m_{h^{\pm}} \gtrsim 300 \text{ GeV}$ \cite{DiLuzio:2015oha}. Perturbativity requirements and EW precision tests lead to rather small mass splittings of $\mathcal{O}$(few GeV) between the neutral and charged components of the triplet. Since we are interested in a non-resonant region of phase space for the Higgs pair production process, we consider scenarios with significantly larger charged Higgs boson masses $m_{h^{\pm}}$ and $m_2$. 
Furthermore, we check for exclusion limits of additional Higgs bosons by means of the code {\tt HiggsBounds} \cite{arXiv:0811.4169, arXiv:1311.0055, arXiv:1507.06706}. It turns out however that for our parameter space scan, no points are excluded. \\
\\
\underline{Perturbativity:}
\\
The adimensional couplings in the potential of \eq{sptriplet} are bounded by perturbative unitarity. 
Looking at correlated matrix of $2 \to 2$ scattering processes one finds \cite{Khan:2016sxm}
\beq 
\label{pertunittriplet}
\lambda_1 < 4 \pi \, , \qquad 
\lambda_2 < 4 \pi \, , \qquad 
\lambda_3 < 8 \pi \, , \qquad 
6 \lambda_1 + 5 \lambda_2 \pm \sqrt{(6 \lambda_1 - 5 \lambda_2)^2 + 12 \lambda_3^2} < 16 \pi \, .
\eeq
For the dimensionful parameter $\mu_4$ we  
estimate the finite loop corrections to the $\mu_4$ vertex at zero external momenta and require it 
to be smaller than the tree-level value. In the $SU(2)_L$ limit we obtain
\beq
\frac{\abs{\mu_4}}{\text{max}\, \left(\abs{\mu_1}, \abs{\mu_2}\right)} < 4 \pi \, .
\eeq
\\
\underline{Vacuum stability:} \\
\\
By requiring that the potential is bounded from below, we obtain the conditions
\bee
\lambda_1 > 0, \hspace{0.5cm}\land\hspace{0.5cm} \lambda_2 > 0, \hspace{0.5cm}\land\hspace*{0.5cm} 
\lambda_3 > - 2 \sqrt{\lambda_1 \lambda_2} \,. \label{eq:boundednesstriplet}
\eee
Also the massive coupling $\mu_4$ can destabilize the potential, if too large. 
We check for vacuum stability using {\tt Vevacious} \cite{Camargo-Molina:2013qva, Camargo-Molina:2014pwa}, 
with a model file generated with {\tt SARAH} \cite{Staub:2008uz,Staub:2009bi,Staub:2010jh,Staub:2012pb,Staub:2013tta}. 

In principle, one should check also for charge breaking (CB) minima. For a CB stationary point we find the necessary condition 
(cf.~\app{app:triplet} for notation)
\bee
\label{stationaryCB}
v^{\eta_+}_{\scriptscriptstyle{CB}} \left( \frac{\lambda_3}{2}v_{\scriptscriptstyle{H,CB}}^2 +\mu_2^2+\lambda_2 v^2_{\scriptscriptstyle{T,CB}} + 2  \lambda_2 |v^{\eta_+}_{\scriptscriptstyle{CB}}|^2\right)=0\,, 
\eee
where the subscript ``CB'' refers to the vevs in the CB minimum and $\langle \eta_+ \eta_-\rangle= |v^{\eta_+}_{\scriptscriptstyle{CB}}|^2$. In addition, from the other stationary equations 
we find that $v_{\scriptscriptstyle{H,CB}}=0$ for $v^{\eta_+}_{\scriptscriptstyle{CB}}\ne 0$ (if $\mu_4 \ne 0$).
Hence, \eq{stationaryCB} implies that non-zero CB stationary points can exists only if
\bee
\frac{1}{2 \lambda_2}\left(\mu_2^2 + \lambda_2 v^2_{\scriptscriptstyle{T,CB}}\right) < 0 \, . 
\label{eq:ccb2}
\eee
Since $\lambda_2 >0$ from the boundedness of the potential, there are no CB stationary points as long as $\mu_2^2>0$. 
We checked explicitly that for all our parameter points $\mu_2^2>0$. 
This can be explained as follows. For $\vt/\vh\ll 1$, we can approximate 
\bee
 \mu_2^2\simeq -\frac{\sin 2 \theta (m_2^2-m_1^2) \vh}{4 \vt} 
 \qquad \text{and} \qquad
 \tan 2 \theta \simeq  \frac{4 \vt}{\vh \mu_4} \left(\lambda_3 \vt - \mu_4 \right) \,.
\eee
Since we work in the basis where $\mu_4 > 0$, the requirement that $m_{h^{\pm}}^2 > 0$ implies $\vt>0$ (cf.~\eq{m2hptriplet}). 
In our scan we use $m_{h^{\pm}}>800\text{ GeV}$. From that we can compute a lower bound on $\mu_4/\vt$ by using Eq.~\eqref{mu4triplet}.
Due to  the perturbativity bound on $\lambda_3$, i.e.~$\lambda_3 < 8\pi /\sqrt{3}$,
from \eq{pertunittriplet} one then finds that $\left(\lambda_3 - \mu_4/\vt \right) <0$. 
Hence, for our scan $\mu_2^2>0$ and we do not need to care for CB minima.

\subsubsection{Results}
\label{sec:results3}
As for the singlet, we perform a scan over the parameter space. 
The scan parameters are 
\bee
m_1=125\text{ GeV}, \hspace*{0.5cm}\;800\text{ GeV}< m_{h^{\pm}} < 4000 \text{ GeV},  \hspace*{0.5cm}v=246.2\text{ GeV},  \; \\ \; \hspace*{-0.4cm} 0<\vt<3.55\text{ GeV}, \hspace*{0.5cm}\; 0.95 < \cos\theta < 1\,,
\hspace*{0.5cm} 0<\lambda_2 < 4\pi\,.
\nonumber
\eee 
It turns out that it is better to scan over $\lambda_2$ rather than $m_2$ 
since the mass difference between $m_2$ and $m_{h^{\pm}}$ is small due to the perturbativity requirement on 
$\lambda_2$ (cf.~\eq{lam2triplet}).
\begin{figure}[h]
\begin{center}
\includegraphics[width=10cm]{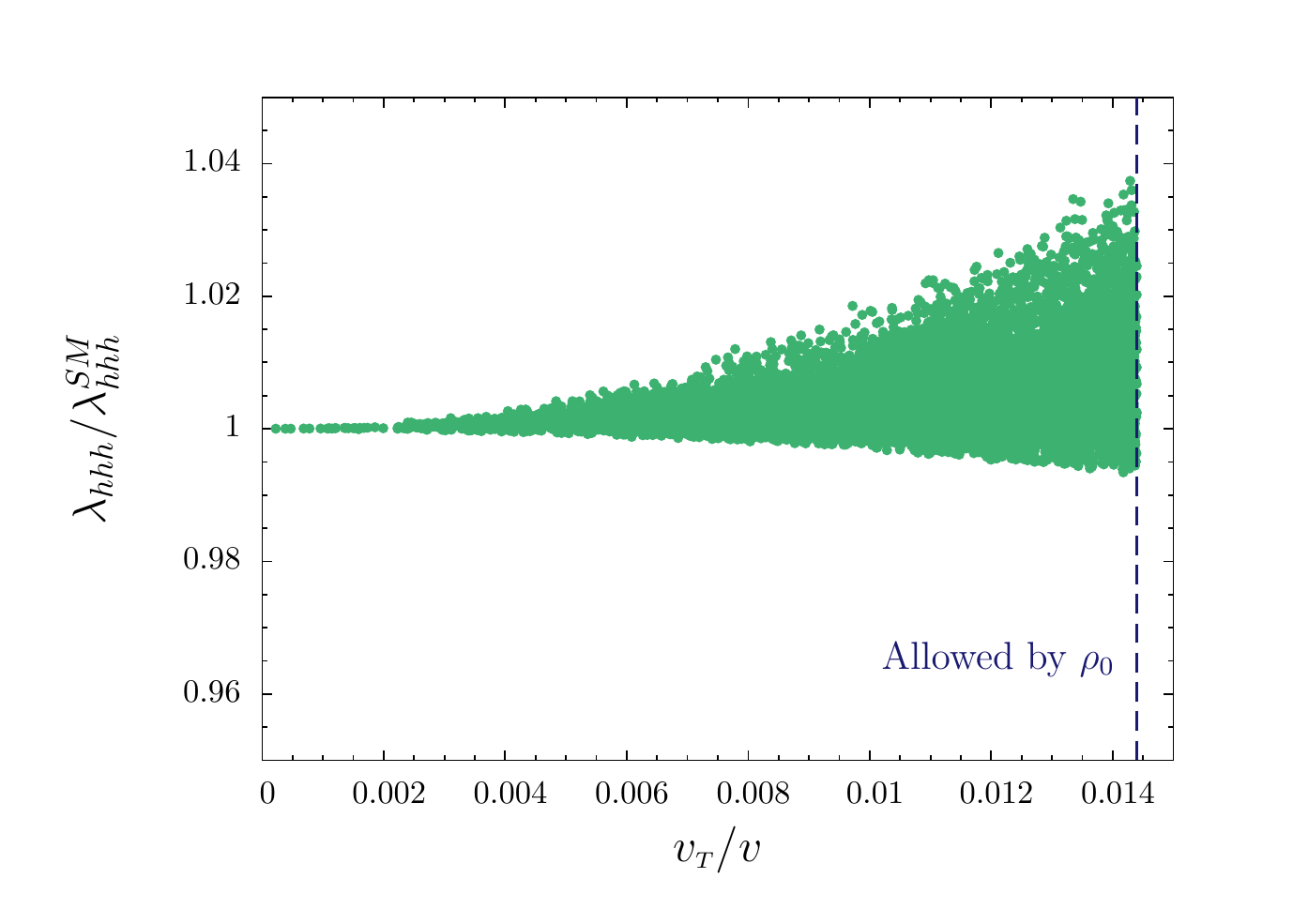}
\caption{The modification of the trilinear Higgs self-coupling with respect to the SM as a function of $\vt/v$. 
For all points the minimum $(\vh, \vt)$ is the global one. \label{fig:triplet}}
\end{center}
\end{figure}
In \fig{fig:triplet} we show the results of our parameter scan. The trilinear Higgs self-coupling can only be modified by a few percent in the triplet model. 
This is a consequence of the small values for $\vt/\vh$ allowed by EW precision data. 

As it can be inferred from the plot all points are stable at tree level. That can be understood as follows.  
In the neutral direction of $H$ the potential has stationary points in $\langle H \rangle=0$ and $\langle H \rangle=\vh/\sqrt{2}$. For $\langle H \rangle=\vh/\sqrt{2}$ the derivative of the potential with respect to the neutral component $\eta^0$ of $\Phi$ reads
\bee
\frac{\partial V}{\partial \eta_0}= \lambda_2 \eta_0^3 + \left( \mu_2^2 +\frac{\lambda_3}{2} \vh^2\right)\eta_0-\frac{\mu_4}{2}\vh^2=0\,.
\eee
The discriminant of the cubic equation then reads
\bee
\Delta=-4 \lambda_2 \left(\mu_2^2+\frac{\lambda_3}{2} \vh^2\right)^2-\frac{27}{4} \lambda_2^2 \,\mu_4^2 \, \vh^4 \,,
\eee
and $\Delta <0$ for all parameter sets due to the boundedness from below condition on $\lambda_2$ from eq.~\eqref{eq:boundednesstriplet}, hence there are no further stationary points with $\langle H \rangle=\vh/\sqrt{2}$ in $H$ direction. Note that for the singlet in \sect{sec:singlet}, due to the $S^3$ term in the potential, the discriminant can also be larger than zero and hence 
other neutral minima can arise. 

Two further stationary points are possible, namely ($\langle H \rangle=0$, $\langle\Phi\rangle=0$) and ($\langle H \rangle=0$, $|\langle \Phi \rangle|^2=-\mu_2^2/\lambda_2$). Since we always find $\mu_2^2>0$ in our scan the latter is not relevant here and  
($\langle H\rangle=0$, $\langle\Phi\rangle=0$) must be a maximum by construction.


It is instructive to compare the previous results with the EFT limit where the triplet mass parameter is $\mu_2 \gg v$. 
By integrating out the triplet in the $SU(2)_L$ limit via the equations of motion 
\beq 
\Phi^\alpha \simeq - \frac{\mu_4}{\mu_2^2 + \lambda_3 \abs{H}^2} H^\dag \sigma^\alpha H 
\eeq 
the potential in the EFT reads 
\beq 
\label{VEFTtriplet}
V_{\rm EFT} (H) \simeq - \frac{1}{2} \frac{\mu_4^2}{\mu_2^2 + \lambda_3 \abs{H}^2} \abs{H}^4 = 
- \frac{\mu_4^2}{2 \mu_2^2} \abs{H}^4 + \frac{\mu_4^2 \lambda_3}{2 \mu_2^4} \abs{H}^6 + \ldots \, ,
\eeq
where the expansion in the last term holds for Higgs fluctuations around the EW vev. 
The first term in \eq{VEFTtriplet} simply redefines the Higgs quartic coupling in the SM EFT, 
while the second one yields 
\beq 
\label{c6triplet}
c_6 = \frac{\mu_4^2 \vh^2 \lambda_3}{2 \mu_2^4} \, .
\eeq 
Always working in the $\mu_2 \gg v$ limit, we can approximate the triplet vev as (cf.~\eq{stateq2trip})
\beq 
\label{vptriplet}
\vt \simeq \frac{\mu_4 \vh^2}{2 \mu_2^2} \, .
\eeq
Hence, it is possible to recast the modified triple Higgs coupling as
\beq 
\frac{\lambda_{hhh}}{\lambda^{\rm SM}_{hhh}} = 1 + \frac{2 c_6 \vh^2 }{m^2_h} =
1 + \frac{4 \vt^2 \lambda_3}{m^2_h} \, ,
\eeq
where in the last step we have replaced $c_6$ in terms of $\vt$ (cf.~\eqs{c6triplet}{vptriplet}). 
By plugging $\vt \lesssim 3.5$ GeV and $\lambda_3 \in [-2 \sqrt{0.1 \times 4 \pi} , 8 \pi / \sqrt{3}]$ from perturbativity and vacuum stability 
(also, $\lambda_1 \sim 0.1$ in order to reproduce the Higgs mass), we get $\lambda_{hhh} / \lambda^{\rm SM}_{hhh} \in [0.99 ,1.046]$, 
which fairly describes the range of deviations in \Fig{fig:triplet}. 

A final comment on the other custodially violating cases is in order. 
By denoting the vev of the complex multiplet as $\vev{\Phi} = v_\Phi / \sqrt{2}$, the 
$2\sigma$-level bound from $\rho_0^{\rm (fit)}$ implies $v_\Phi \lesssim 1.7$, $2.9$, $1.0$ GeV, respectively for 
$\Phi = (1,3,1), (1,4, \frac{1}{2}), (1,4, \frac{3}{2})$. 
We hence expect suppressed contributions for the trilinear Higgs self-coupling, similarly to the triplet case.

\subsection{Loop-induced trilinear Higgs self-coupling vs.~vacuum stability}
\label{looptrilinear}

\begin{figure}[h!]
\centering
\includegraphics[angle=0,width=10.cm]{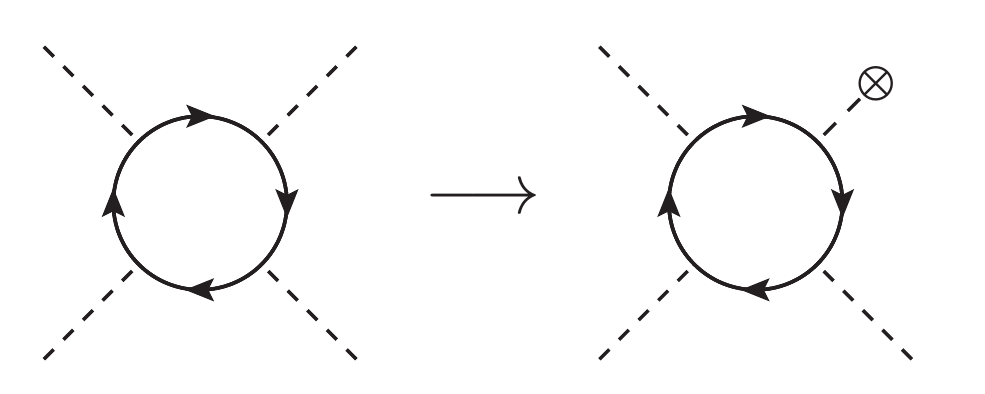}
\caption{\label{fig:betalamvshhh} 
Schematic view of the connection between the beta-function of $\lambda$ and the loop-induced 
trilinear Higgs self-coupling via new fermions.  
}
\end{figure}

Loop modifications of the trilinear Higgs self-coupling are naturally expected to be smaller than tree-level ones. 
Nevertheless, we consider here the case where the new particles circulating in the loops are vector-like fermions, 
since we regain a clean correlation between the triple Higgs coupling and vacuum 
instability. This can be easily understood by looking at the loop of fermions 
contributing to the beta function of the Higgs self-coupling, 
which is basically the same diagram responsible for the radiative generation of the 
trilinear Higgs self-coupling in the broken phase after taking one Higgs to its vev (cf.~\fig{fig:betalamvshhh}).
 
There are basically two qualitatively different possibilities: $i)$ non-SM-singlet fermions 
coupling to the Higgs and a SM fermion and $ii)$ SM-singlet fermions coupling to the Higgs and a lepton 
doublet. The former cases are bounded by other Higgs coupling measurements, which typically 
imply a very suppressed contribution to the trilinear Higgs self-coupling. 
The latter is more interesting, and correspond to the case of a 
right handed neutrino, which is largely unconstrained by other Higgs coupling measurements. 
A recent analysis was performed in Refs.~\cite{Baglio:2016ijw,Baglio:2016bop} in the context 
of a simplified $3+1$ Dirac neutrino model \cite{Baglio:2016ijw} and for the inverse seesaw model \cite{Baglio:2016bop}, 
finding deviations of the trilinear Higgs self-coupling with respect to the SM value up to $30 \%$. 

We want to show here the impact of vacuum stability in such a class of scenarios. 
Let us consider, for definiteness, the case of the inverse seesaw (similar conclusions apply to other neutrino mass models as well). 
We add to the SM field content three right-handed neutrinos and three gauge singlets $X$ with opposite lepton number, 
via the Lagrangian term 
\beq 
\label{LISS}
\mathcal{L}_{\rm ISS} = - Y_\nu \overline{L} \tilde H \nu_R - M_R \overline{\nu^c} X - \frac{1}{2} \mu_X \overline{X^c} X + \text{h.c.} \, ,
\eeq
where $\tilde H = i \sigma_2 H^*$ and we suppressed family indices. We refer to Ref.~\cite{Baglio:2016bop} for the relevant notation and conventions. 
Taking, in particular, a diagonal Yukawa structure $Y_\nu = \abs{y_\nu} I_{3}$ and a common mass scale for the three heavy neutrinos, $M_R = 10$ TeV, 
one can asses the impact of the heavy neutrino states on the running of the Higgs self-coupling and hence on the stability of the 
Higgs effective potential $V_{\rm eff} (h) \approx 1/4 \, \lambda_{\rm eff} (h) h^4$, where $\lambda_{\rm eff} (h)$ is approximated with the 
$\overline{\text{MS}}$ running coupling $\lambda (\mu = h)$. 
We use the two-loop beta functions for the SM couplings $(g_{1,2,3}, y_t, \lambda)$ and take into account the corrections due to 
$y_\nu$ at the one-loop level (and consistently we neglect the matching contributions of $y_\nu$ to $\lambda (M_t)$).  
For simplicity, we also integrate in the heavy neutrinos at the common threshold $M_R = 10$ TeV, 
while a more careful treatment should take into account intermediate EFTs when integrating in 
single neutrino thresholds (see e.g.~Ref.~\cite{Rose:2015fua}). Hence, in the case of a hierarchical heavy neutrino spectrum, 
our estimate of the largest energy scale until which the model can be consistently extrapolated 
should be conservatively rescaled starting from the heaviest threshold.  

\begin{figure}[h!]
\centering
\includegraphics[angle=0,width=7.cm]{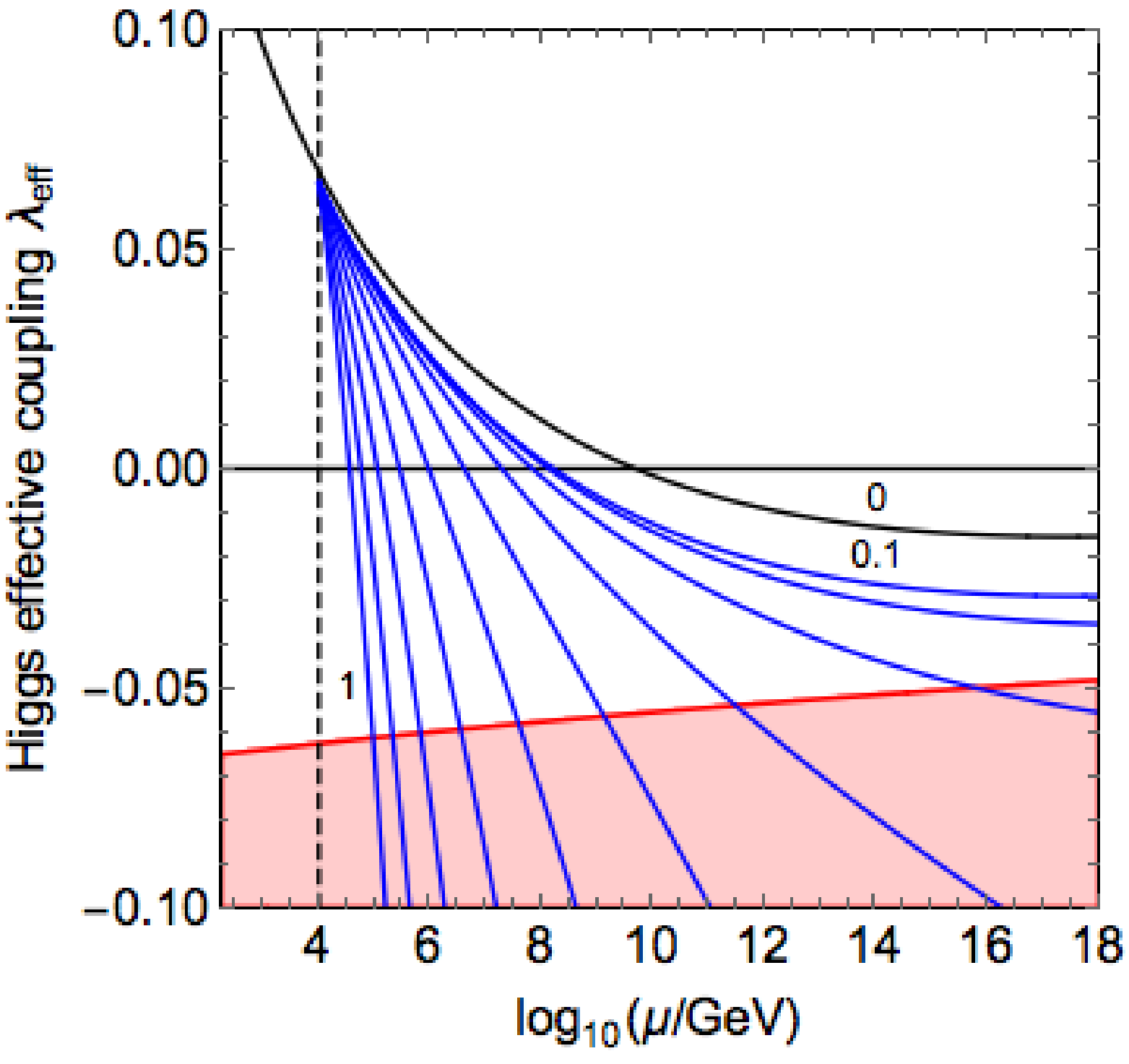}
\caption{\label{fig:Instability_loopinduced} 
Running of $\lambda_{\rm eff}$ in the presence of a common heavy neutrino threshold $M_R = 10$ TeV. 
Labels denote the the value of $y_\nu \in [0.1, 1]$ with steps of $0.1$ (blue curves), 
while $y_\nu = 0$ corresponds to the SM case (black curve). The instability bound is represented by the red-shaded area. 
}
\end{figure}

The results are displayed in \fig{fig:Instability_loopinduced} where we plot the value of $\lambda_{\rm eff}$ 
as a function of the renormalization scale $\mu$. The instability bound (red area) is computed by 
considering the probability of decay against quantum tunnelling in the modified Higgs potential 
integrated over the past light-cone (see e.g.~\cite{Isidori:2001bm,DiLuzio:2015iua})
\beq 
\mathcal{P}_{\rm{EW}} \simeq \left(\frac{\mu}{H_0}\right)^4 e^{-\frac{8 \pi^2}{3 \abs{\lambda_{\rm{eff}}(\mu)}}} \, ,
\eeq
where $H_0 \simeq 10^{-42}$ GeV is the present Hubble constant. In particular, requiring $\mathcal{P}_{\rm{EW}} \simeq 1$ 
corresponds to 
\beq 
\abs{\lambda_{\rm{eff}}(\mu)} 
\simeq \frac{0.064}{1 + 0.022 \, \log_{10} \left(  \frac{\mu}{1 \ \rm{TeV}} \right)} \, , 
\eeq
which sets the instability bound for $\lambda_{\rm eff} < 0$. 

By increasing the value of $y_\nu$ between $0.1$ and $1$ (in steps of $0.1$),  
the instability scale dangerously approaches the heavy neutrino threshold (see \fig{fig:Instability_loopinduced}),   
and in order to comply with the existence of the EW vacuum the 
model must be UV completed before entering the instability region. 
Using the approximate expression for $\Delta^{\rm BSM}_{\rm approx} \equiv \lambda_{hhh} / \lambda^{\rm SM}_{hhh} - 1$ in 
Eq.~(4.5) of \cite{Baglio:2016bop} we obtain that $y_\nu = 0.8$ corresponds to $\Delta^{\rm BSM}_{\rm approx} = 0.1 \ \%$. 
Hence, from \fig{fig:Instability_loopinduced} we read that modifications of the trilinear Higgs self-coupling 
above the per mil level require an UV completion within a few orders of magnitude 
from the scale where the heavy neutrinos are integrated in. 


\section{Conclusions}
\label{sec:conclusions}

In this paper we have addressed the question on how much could the 
trilinear Higgs self-coupling deviate from its SM value. 
We first discussed in \sect{sec:EFT} theoretical constraints on Higgs self-couplings from a general standpoint  
by considering two main arguments: vacuum instability and pertubativity. 
We showed that the former cannot be reliably assessed in a model-independent way, 
due to the breakdown of the EFT in describing the global structure of the Higgs potential 
away from the EW minimum. In particular, we have explicitly shown that by augmenting the SM via an $\abs{H}^6$ 
operator one can generate two type of instabilities, either at large field values $v \ll H \lesssim \Lambda$ or in $H = 0$. 
In both cases, however, any reliable statement about the stability of the EW vacuum 
entails the knowledge of the full tower of effective operators, thus jeopardizing the connection with the 
Higgs self-couplings, whose leading order deviations are still governed by the $d=6$ operators.

On the other hand, it is possible to use perturbativity in order to set fairly model-independent limits on 
Higgs self-couplings. In \sect{sectperthhh} we have employed two different criteria, 
based either on the partial-wave unitarity of the $hh \to hh $ scattering 
or on the loop corrections of the tree-level vertices, 
in order to establish the perturbative domain of the Higgs self-couplings. 
Though being conceptually different, 
the two criteria agree well with each other both for the triple and the quartic Higgs coupling modifications: 
$|\lambda_{hhh}/\lambda_{hhh}^{SM}| \lesssim 6.5 \, (6.0)$ and 
$|\lambda_{hhhh}/\lambda_{hhhh}^{SM}| \lesssim 65 \, (68)$, 
with the first number corresponding to perturbative unitarity and 
the one in the bracket stemming from the loop-corrected vertex. 
Let us stress that indirect tests of the trilinear Higgs self-coupling 
either via single Higgs production \cite{Gorbahn:2016uoy,Degrassi:2016wml,Bizon:2016wgr} 
or EW precision tests \cite{Kribs:2017znd,Degrassi:2017ucl} 
and current measurements of non-resonant Higgs pair production \cite{CMS-PAS-HIG-17-006}
bound values of $\lambda_{hhh}$ which are, at the moment, well above our 
perturbativity limit $\abs{\lambda_{hhh} / \lambda^{\rm SM}_{hhh}} \lesssim 6$.

In the second part of the paper (\sect{UVmodels}), we investigated the size of the trilinear Higgs self-coupling 
in explicit models. First, we identified the class of models potentially leading to the 
largest modifications in the trilinear Higgs self-coupling, namely scalar extensions featuring a tadpole operator 
of the type $\mathcal{O}_\Phi = \Phi f(H)$, 
where $f(H)$ is a string of Higgs fields. The list of new scalars coupling linearly to $H$ can be found 
in \Table{newscalarsHHH}. They include both custodial symmetric (EW singlet and doublet) and custodial violating (EW triplets and quadruplets) scalar extensions. As two representative examples, we studied in detail the size of the trilinear Higgs self-coupling in the 
singlet and triplet extension, by taking into account constraints from EW precision tests, Higgs coupling measurements, 
direct searches for new scalars, vacuum stability and perturbativity. 
While in the singlet extension modifications of the trilinear Higgs coupling 
in the range $-1.5 < \lambda_{hhh}/\lambda_{hhh}^{\rm SM} < 8.7$ 
are still possible,
for the custodially violating extensions, like e.g.~the triplet case, 
only modifications up to few percent are allowed.

Remarkably, vacuum stability is not a crucial discriminant for limiting the size of the trilinear Higgs self-coupling 
in models featuring new scalars, where the intricate structure of the scalar potential allows for 
regions in parameter space where large quartics (at the boundary of perturbativity) can tame the instabilities triggered 
by the tadpole operators. On the other hand, we have also found circumstances where vacuum stability 
can be very relevant. That is the case 
in which the trilinear Higgs self-coupling is modified by loops of heavy fermions. 
In our explicit example in \sect{looptrilinear} we have considered the case of low-scale seesaw models, 
where the vacuum metastability bound can sizably reduce the allowed range for the trilinear Higgs self-coupling. 

\vskip 1 \baselineskip

\noindent {\it{Acknowledgments. We thank Julien Baglio, Marco Nardecchia, Enrico Nardi, Michael Spira and 
C\'edric Weiland for helpful discussions. RG is supported by a European Union COFUND/Durham Junior Research Fellowship under the EU grant number 609412. MS is supported in part by the European Commission through the ``HiggsTools'' Initial Training Network PITN-GA-2012-316704.}}

  
\begin{appendix}
  
\section{Scalar potential parameters} 

In this appendix we collect some details on the scalar potential (e.g.~tadpole equations and scalar spectrum)
for the two models studied in \sect{UVmodels}. 

\subsection{Singlet}
\label{app:singlet}

The scalar fields can be expanded around their vevs by
\begin{equation}
H=\frac{1}{\sqrt{2}} \left( \begin{array}{c} 0 \\ \vh +h \end{array} \right) \, ,  
\qquad 
\Phi= (\vs+S)
\,, 
\end{equation}
where we employed the unitary gauge for the Higgs doublet. 
The tadpole conditions can be written as
\begin{align}
-\mu_4 \vs-\frac{\lambda_3 \vs^2}{2}-\mu_1^2-\lambdah
  \vh^2= 0 \,,\\
 -\frac{\mu_4 \vh^2}{2 }-\frac{1}{2} \lambda_3 \vh^2 \vs-\mu_2^2
  \vs-\lambda_{2}  \vs^3-\mu_3  \vs^2= 0 \,. \label{eq:tadpole}
\end{align}
The first condition allows to replace $\mu_1^2$ in terms of $\vh$.
The mass matrix in the real $(h,S)$ basis then reads
\begin{equation}
\mathcal{M}_0^2 =
\left(\begin{array}{cc} m_{\scriptscriptstyle{hh}} & m_{\scriptscriptstyle{hS}} \\
m_{\scriptscriptstyle{hS}} & m_{\scriptscriptstyle{SS}}
 \end{array}\right) \, ,
\end{equation}
with
\begin{align}
m_{\scriptscriptstyle{hh}}&= 2 \vh^2 \lambdah \,,\\
m_{\scriptscriptstyle{hS}}&=\vh \left(\mu_4 + \lambda_3 \vs \right)\,,\\
m_{\scriptscriptstyle{SS}}&= \mu_2^2 +\frac{1}{2} \left(\lambda_3 \vh^2 +6 \vs^2 \lambda_{2}+4 \vs \mu_3 \right)\,.
\end{align}
The mass matrix is diagonalized by rotating
\begin{equation}
\left(\begin{array}{c} h_1 \\ h_2 \end{array}\right)=\left( \begin{array}{cc} \cos\theta &- \sin\theta \\ \sin\theta & \cos\theta \end{array}\right) \left(\begin{array}{c} h \\ S \end{array}\right)\,,
\label{eq:mixingangle}
\end{equation}
with 
\beq 
\tan 2 \theta =\frac{2 m_{\scriptscriptstyle{hS}}}{m_{\scriptscriptstyle{SS}}-m_{\scriptscriptstyle{hh}}} \, ,
\eeq 
and mass eigenvalues
\begin{align}
m_{1,2}^2&=\frac{1}{2}\left(m_{\scriptscriptstyle{hh}}+m_{\scriptscriptstyle{SS}} \mp \sqrt{4 m_{\scriptscriptstyle{hS}}^2 +(m_{\scriptscriptstyle{hh}}-m_{\scriptscriptstyle{SS}})^2}\right) \nonumber \\
&= \frac{1}{2}\left( m_{\scriptscriptstyle{hh}}+ m_{\scriptscriptstyle{SS}} \pm (m_{\scriptscriptstyle{hh}}-m_{\scriptscriptstyle{SS}}) \frac{1}{\cos 2 \theta}\right) \, .
\end{align}
Expressing the couplings of the potential in terms of the parameters used for the scan, we find 
\begin{align}
\mu_1^2 &=- \frac{1}{4}\left[\left(-2 \lambda_3 \vs^2+m_1^2+m_2^2 \right)+ \cos (2 \theta ) (m_1^2-m_2^2)
- 2 \frac{\vs}{\vh}\sin (2 \theta )
   (m_1^2-m_2^2) \right]\,,  \label{eq:muh}\\
    \mu_2^2 &=\frac{1}{2}\left[\left( \lambda_3 \vh^2- m_1^2- m_2^2+2\lambda_{2 }
   \vs^2\right)+\frac{\vh}{\vs} \sin (2 \theta ) (m_1^2-m_2^2)+
   \cos (2 \theta ) (m_1^2-m_2^2)\right]\,, \label{eq:m2} \\
   \mu_3 &= \frac{1} {2 \vs}\left[ \left( m_1^2+
   m_2^2- \lambda_3 \vh^2 -4\lambda_{2}  \vs^2\right)-\frac{1}{2} \frac{\vh}{\vs} \sin (2 \theta )
   (m_1^2-m_2^2)- \cos (2 \theta )
   (m_1^2-m_2^2)\right]\,, \label{eq:musinglet}\\
   \mu_4& = \frac{\sin (2 \theta ) (m_2^2-m_1^2)-2 \lambda_3 \vh \vs}{2
   \vh}  \,, \\
   \lambdah &= \frac{\cos (2 \theta ) (m_1^2-m_2^2)+m_1^2+m_2^2}{4
   \vh^2}\,.  \label{eq:lambdah}\\
\end{align}

\subsection{Triplet}
\label{app:triplet}

The scalar fields can be expanded around their charge-preserving vevs 
via
\begin{equation}
H= \left( \begin{array}{c} 
\phi^+ \\ 
\frac{1}{\sqrt{2}} \left( \vh +h^0 + i G^0 \right) \end{array} \right) \, , \hspace*{1cm}  
\Phi= \left( \begin{array}{c} 
\eta_1 \\
\eta_2 \\ 
\vt + \eta_0 \end{array} \right) \,,
\end{equation}
where the charged eigenstates for $\Phi$ are defined as $\eta^{\pm} = \tfrac{1}{\sqrt{2}} \left( \eta_1 \mp i \eta_2 \right)$. 
The tadpole conditions can be written as
\begin{align}
\label{stateq1trip}
0 &= \mu_1^2 + \lambda_1 \vh^2 + \frac{\lambda_3}{2} \vt^2 - \mu_4 \vt \, , \\
\label{stateq2trip}
0 &= \vt \left( \mu_2^2 + \lambda_2 \vt^2 + \frac{\lambda_3}{2} \vh^2 \right) - \frac{\mu_4}{2} \vh^2 \, . 
\end{align}
For $\vt = 0$ there is no doublet/triplet mixing and \eq{stateq2trip} implies $\mu_4 = 0$, 
which corresponds to the custodial symmetric tree-level relation $\rho=1$. From now on we will assume $\vt \neq 0$. 
By evaluating the second derivatives of the scalar potential and after imposing the stationary \eqs{stateq1trip}{stateq1trip}, 
we find the following scalar spectrum: 
\begin{itemize}
\item Charged scalars: in the complex $(\phi^+, \eta^+)$ basis
\beq 
\mathcal{M}^2_+ = 
\left( 
\begin{array}{cc}
2 \mu_4 \vt & \mu_4 \vh \\
\mu_4 \vh & \frac{\mu_4 \vh^2}{2 \vt} 
\end{array}
\right) \, , 
\eeq 
which features a null eigenvalue, corresponding to the Goldstone boson $G^+$ eaten by the $W$, 
and a massive state $h^\pm$ with mass 
\beq 
\label{m2hptriplet}
m^2_{h^\pm} = \frac{\mu_4 \left( \vh^2 + 4 \vt^2 \right)}{2 \vt} \, . 
\eeq
\item Neutral pseudo-scalar: $G^0$, corresponding to the Goldstone boson eaten by the $Z$. 
\item Neutral scalars: in the real $(h^0, \eta^0)$ basis 
\begin{equation}
\mathcal{M}_0^2 =
\left(\begin{array}{cc} m_{\scriptscriptstyle{hh}} & m_{\scriptscriptstyle{h\eta}} \\
m_{\scriptscriptstyle{h\eta}} & m_{\scriptscriptstyle{\eta\eta}}
 \end{array}\right) \, ,
\end{equation}
with
\begin{align}
m_{\scriptscriptstyle{hh}}&= 2 \lambda_1 \vh^2 \,,\\
m_{\scriptscriptstyle{h\eta}}&=v \left(\lambda_3 \vt - \mu_4 \right)\,,\\
m_{\scriptscriptstyle{\eta\eta}}&= 2 \lambda_2 \vt^2 + \frac{\mu_4 \vh^2}{2 \vt}\,.
\end{align}
The mass eigenstates are obtained via the rotation 
\begin{equation}
\left(\begin{array}{c} h_1 \\ h_2 \end{array}\right)=\left( \begin{array}{cc} \cos\theta &- \sin\theta \\ \sin\theta & \cos\theta \end{array}\right) \left(\begin{array}{c} h^0 \\ \eta^0 
\end{array}\right)\,,
\end{equation}
with 
\beq 
\label{tan2thtriplet}
\tan 2 \theta =\frac{2 \vh \left(\lambda_3 \vt - \mu_4 \right)}{2 \lambda_2 \vt^2 + \frac{\mu_4 \vh^2}{2 \vt} - 2 \lambda_1 \vh^2} \,, 
\eeq
and mass eigenvalues
\begin{align}
m_{1,2}^2
&=\frac{1}{2}\left(m_{\scriptscriptstyle{hh}}+m_{\scriptscriptstyle{\eta\eta}} \mp \sqrt{4 m_{\scriptscriptstyle{h\eta}}^2 +(m_{\scriptscriptstyle{hh}}-m_{\scriptscriptstyle{\eta\eta}})^2}\right) 
\nonumber \\
&= \frac{1}{2}\left( m_{\scriptscriptstyle{hh}}+ m_{\scriptscriptstyle{\eta\eta}} \pm (m_{\scriptscriptstyle{hh}}-m_{\scriptscriptstyle{\eta\eta}}) \frac{1}{\cos 2 \theta}\right) \, . 
\end{align}
\end{itemize}
Moreover, the $W$ boson mass is given by 
\beq 
m^2_W = \frac{g^2}{4} \left( \vh^2 + 4 \vt^2 \right) \, , 
\eeq
which fixes $v^2 = (246.2 \ \text{GeV})^2 = \vh^2 + 4 \vt^2$, while EW precision tests set a bound on the custodial-breaking vev $\vt \lesssim 3.5$ GeV. 
Summarising, an independent set of parameters can be chosen as: 
\beeq
\label{scanpartriplet}
\vh = \sqrt{v^2 - 4 \vt^2} \,, \; \; \; \vt < 3.55 \ \text{GeV} \,, \; \; \; m_1=125 \text{ GeV}\,, \; \; \; m_2\,, \; \; \; m_{h^\pm}\,, \; \; \;\theta \,.
\eeeq 
Note, however, that in \sect{sec:triplet} we scan over $\lambda_2$ instead of $m_2$.
For completeness, we report here the potential parameters expressed in terms of those in \eq{scanpartriplet}
\begin{align}
\label{mu4triplet}
\mu_4 &= \frac{2 m^2_{h^\pm} \vt}{\vh^2 + 4 \vt^2} \, , \\
\label{lam1triplet}
\lambda_1 &= \frac{m_1^2 + m_2^2 + (m_1^2 - m_2^2) \cos 2\theta}{4 \vh^2} \, , \\
\label{lam2triplet}
\lambda_2 &= \frac{(m_1^2 + m_2^2) \vt - \mu_4 \vh^2 + (m_2^2 - m_1^2) \vt \cos 2\theta}{4 \vt^3} \, , \\
\label{lam3triplet}
\lambda_3 &= \frac{2 \mu_4 \vh + (m_2^2 - m_1^2) \sin 2\theta}{2 \vh \vt} \, .
\end{align}

\end{appendix}
  
\bibliographystyle{utphys.bst}
\bibliography{references}

\providecommand{\href}[2]{#2}\begingroup\raggedright\begin{thebibliography}{100}

\bibitem{Aad:2012tfa}
{\bfseries ATLAS} Collaboration, G.~Aad {\em et~al.}, ``{Observation of a new
  particle in the search for the Standard Model Higgs boson with the ATLAS
  detector at the LHC},''
  \href{http://dx.doi.org/10.1016/j.physletb.2012.08.020}{{\em Phys. Lett.}
  {\bfseries B716} (2012) 1--29},
\href{http://arxiv.org/abs/1207.7214}{{\ttfamily arXiv:1207.7214 [hep-ex]}}.

\bibitem{Chatrchyan:2012xdj}
{\bfseries CMS} Collaboration, S.~Chatrchyan {\em et~al.}, ``{Observation of a
  new boson at a mass of 125 GeV with the CMS experiment at the LHC},''
  \href{http://dx.doi.org/10.1016/j.physletb.2012.08.021}{{\em Phys. Lett.}
  {\bfseries B716} (2012) 30--61},
\href{http://arxiv.org/abs/1207.7235}{{\ttfamily arXiv:1207.7235 [hep-ex]}}.

\bibitem{Englert:1964et}
F.~Englert and R.~Brout, ``{Broken Symmetry and the Mass of Gauge Vector
  Mesons},''
\href{http://dx.doi.org/10.1103/PhysRevLett.13.321}{{\em Phys. Rev. Lett.}
  {\bfseries 13} (1964) 321--323}.

\bibitem{Higgs:1964pj}
P.~W. Higgs, ``{Broken Symmetries and the Masses of Gauge Bosons},''
\href{http://dx.doi.org/10.1103/PhysRevLett.13.508}{{\em Phys. Rev. Lett.}
  {\bfseries 13} (1964) 508--509}.

\bibitem{Klinkhamer:1984di}
F.~R. Klinkhamer and N.~S. Manton, ``{A Saddle Point Solution in the
  Weinberg-Salam Theory},''
\href{http://dx.doi.org/10.1103/PhysRevD.30.2212}{{\em Phys. Rev.} {\bfseries
  D30} (1984) 2212}.

\bibitem{Spannowsky:2016ile}
M.~Spannowsky and C.~Tamarit, ``{Sphalerons in composite and non-standard Higgs
  models},'' \href{http://dx.doi.org/10.1103/PhysRevD.95.015006}{{\em Phys.
  Rev.} {\bfseries D95} no.~1, (2017) 015006},
\href{http://arxiv.org/abs/1611.05466}{{\ttfamily arXiv:1611.05466 [hep-ph]}}.

\bibitem{ATLAS-CONF-2016-004}
{\bfseries ATLAS} Collaboration, ``{Search for Higgs boson pair production in
  the $b\bar{b}\gamma\gamma$ final state using pp collision data at
  $\sqrt{s}=13$ TeV with the ATLAS detector},'' Tech. Rep. ATLAS-CONF-2016-004,
  CERN, Geneva, Mar, 2016.
\newblock \url{http://cds.cern.ch/record/2138949}.

\bibitem{Aaboud:2016xco}
{\bfseries ATLAS} Collaboration, M.~Aaboud {\em et~al.}, ``{Search for pair
  production of Higgs bosons in the $b\bar{b}b\bar{b}$ final state using
  proton--proton collisions at $\sqrt{s} = 13$ TeV with the ATLAS detector},''
  \href{http://dx.doi.org/10.1103/PhysRevD.94.052002}{{\em Phys. Rev.}
  {\bfseries D94} no.~5, (2016) 052002},
\href{http://arxiv.org/abs/1606.04782}{{\ttfamily arXiv:1606.04782 [hep-ex]}}.

\bibitem{ATLAS-CONF-2016-071}
{\bfseries ATLAS} Collaboration, ``{Search for Higgs boson pair production in
  the final state of $\gamma\gamma WW^*$($\rightarrow l\nu jj$) using 13.3
  fb$^{-1}$ of $pp$ collision data recorded at $\sqrt{s}= $ 13 TeV with the
  ATLAS detector},'' Tech. Rep. ATLAS-CONF-2016-071, CERN, Geneva, Aug, 2016.
\newblock \url{http://cds.cern.ch/record/2206222}.

\bibitem{CMS-PAS-HIG-16-026}
{\bfseries CMS} Collaboration, ``{Search for non-resonant pair production of
  Higgs bosons in the $\rm{b} \bar{\rm{b}} \rm{b} \bar{\rm{b}}$ final state
  with 13 TeV CMS data},'' Tech. Rep. CMS-PAS-HIG-16-026, CERN, Geneva, 2016.
\newblock \url{https://cds.cern.ch/record/2209572}.

\bibitem{CMS-PAS-HIG-17-002}
{\bfseries CMS} Collaboration, ``{Search for pair production of Higgs bosons in
  the two tau leptons and two bottom quarks final state using proton-proton
  collisions at $\sqrt{s} = 13~\mathrm{TeV}$},'' Tech. Rep. CMS-PAS-HIG-17-002,
  CERN, Geneva, 2017.
\newblock \url{https://cds.cern.ch/record/2256096}.

\bibitem{CMS-PAS-HIG-17-006}
{\bfseries CMS} Collaboration, ``{Search for resonant and non-resonant Higgs
  boson pair production in the $\mathrm{b}\overline{\mathrm{b}} \mathrm{l}\nu
  \mathrm{l}\nu$ final state at $\sqrt{s} = 13~\mathrm{TeV}$},'' Tech. Rep.
  CMS-PAS-HIG-17-006, CERN, Geneva, 2017.
\newblock \url{https://cds.cern.ch/record/2257068}.

\bibitem{Djouadi:1999gv}
A.~Djouadi, W.~Kilian, M.~Muhlleitner, and P.~M. Zerwas, ``{Testing Higgs
  selfcouplings at e+ e- linear colliders},''
  \href{http://dx.doi.org/10.1007/s100529900082}{{\em Eur. Phys. J.} {\bfseries
  C10} (1999) 27--43},
\href{http://arxiv.org/abs/hep-ph/9903229}{{\ttfamily arXiv:hep-ph/9903229
  [hep-ph]}}.

\bibitem{Djouadi:1999rca}
A.~Djouadi, W.~Kilian, M.~Muhlleitner, and P.~M. Zerwas, ``{Production of
  neutral Higgs boson pairs at LHC},''
  \href{http://dx.doi.org/10.1007/s100529900083}{{\em Eur. Phys. J.} {\bfseries
  C10} (1999) 45--49},
\href{http://arxiv.org/abs/hep-ph/9904287}{{\ttfamily arXiv:hep-ph/9904287
  [hep-ph]}}.

\bibitem{Baur:2002qd}
U.~Baur, T.~Plehn, and D.~L. Rainwater, ``{Determining the Higgs boson
  selfcoupling at hadron colliders},''
  \href{http://dx.doi.org/10.1103/PhysRevD.67.033003}{{\em Phys. Rev.}
  {\bfseries D67} (2003) 033003},
\href{http://arxiv.org/abs/hep-ph/0211224}{{\ttfamily arXiv:hep-ph/0211224
  [hep-ph]}}.

\bibitem{Baur:2003gp}
U.~Baur, T.~Plehn, and D.~L. Rainwater, ``{Probing the Higgs selfcoupling at
  hadron colliders using rare decays},''
  \href{http://dx.doi.org/10.1103/PhysRevD.69.053004}{{\em Phys. Rev.}
  {\bfseries D69} (2004) 053004},
\href{http://arxiv.org/abs/hep-ph/0310056}{{\ttfamily arXiv:hep-ph/0310056
  [hep-ph]}}.

\bibitem{Dolan:2012rv}
M.~J. Dolan, C.~Englert, and M.~Spannowsky, ``{Higgs self-coupling measurements
  at the LHC},'' \href{http://dx.doi.org/10.1007/JHEP10(2012)112}{{\em JHEP}
  {\bfseries 10} (2012) 112},
\href{http://arxiv.org/abs/1206.5001}{{\ttfamily arXiv:1206.5001 [hep-ph]}}.

\bibitem{Baglio:2012np}
J.~Baglio, A.~Djouadi, R.~Gr{\"o}ber, M.~M. M{\"u}hlleitner, J.~Quevillon, and
  M.~Spira, ``{The measurement of the Higgs self-coupling at the LHC:
  theoretical status},'' \href{http://dx.doi.org/10.1007/JHEP04(2013)151}{{\em
  JHEP} {\bfseries 04} (2013) 151},
\href{http://arxiv.org/abs/1212.5581}{{\ttfamily arXiv:1212.5581 [hep-ph]}}.

\bibitem{Barr:2013tda}
A.~J. Barr, M.~J. Dolan, C.~Englert, and M.~Spannowsky, ``{Di-Higgs final
  states augMT2ed -- selecting $hh$ events at the high luminosity LHC},''
  \href{http://dx.doi.org/10.1016/j.physletb.2013.12.011}{{\em Phys. Lett.}
  {\bfseries B728} (2014) 308--313},
\href{http://arxiv.org/abs/1309.6318}{{\ttfamily arXiv:1309.6318 [hep-ph]}}.

\bibitem{Dolan:2013rja}
M.~J. Dolan, C.~Englert, N.~Greiner, and M.~Spannowsky, ``{Further on up the
  road: $hhjj$ production at the LHC},''
  \href{http://dx.doi.org/10.1103/PhysRevLett.112.101802}{{\em Phys. Rev.
  Lett.} {\bfseries 112} (2014) 101802},
\href{http://arxiv.org/abs/1310.1084}{{\ttfamily arXiv:1310.1084 [hep-ph]}}.

\bibitem{Papaefstathiou:2012qe}
A.~Papaefstathiou, L.~L. Yang, and J.~Zurita, ``{Higgs boson pair production at
  the LHC in the $b \bar{b} W^+ W^-$ channel},''
  \href{http://dx.doi.org/10.1103/PhysRevD.87.011301}{{\em Phys. Rev.}
  {\bfseries D87} no.~1, (2013) 011301},
\href{http://arxiv.org/abs/1209.1489}{{\ttfamily arXiv:1209.1489 [hep-ph]}}.

\bibitem{Goertz:2013kp}
F.~Goertz, A.~Papaefstathiou, L.~L. Yang, and J.~Zurita, ``{Higgs Boson
  self-coupling measurements using ratios of cross sections},''
  \href{http://dx.doi.org/10.1007/JHEP06(2013)016}{{\em JHEP} {\bfseries 06}
  (2013) 016},
\href{http://arxiv.org/abs/1301.3492}{{\ttfamily arXiv:1301.3492 [hep-ph]}}.

\bibitem{deLima:2014dta}
D.~E. Ferreira~de Lima, A.~Papaefstathiou, and M.~Spannowsky, ``{Standard model
  Higgs boson pair production in the ( $ b\overline{b} $ )( $ b\overline{b} $ )
  final state},'' \href{http://dx.doi.org/10.1007/JHEP08(2014)030}{{\em JHEP}
  {\bfseries 08} (2014) 030},
\href{http://arxiv.org/abs/1404.7139}{{\ttfamily arXiv:1404.7139 [hep-ph]}}.

\bibitem{Englert:2014uqa}
C.~Englert, F.~Krauss, M.~Spannowsky, and J.~Thompson, ``{Di-Higgs
  phenomenology in $t\bar{t}hh$: The forgotten channel},''
  \href{http://dx.doi.org/10.1016/j.physletb.2015.02.041}{{\em Phys. Lett.}
  {\bfseries B743} (2015) 93--97},
\href{http://arxiv.org/abs/1409.8074}{{\ttfamily arXiv:1409.8074 [hep-ph]}}.

\bibitem{Liu:2014rva}
T.~Liu and H.~Zhang, ``{Measuring Di-Higgs Physics via the $t \bar t hh \to t
  \bar t b \bar bb\bar b$ Channel},''
\href{http://arxiv.org/abs/1410.1855}{{\ttfamily arXiv:1410.1855 [hep-ph]}}.

\bibitem{Goertz:2014qta}
F.~Goertz, A.~Papaefstathiou, L.~L. Yang, and J.~Zurita, ``{Higgs boson pair
  production in the D=6 extension of the SM},''
  \href{http://dx.doi.org/10.1007/JHEP04(2015)167}{{\em JHEP} {\bfseries 04}
  (2015) 167},
\href{http://arxiv.org/abs/1410.3471}{{\ttfamily arXiv:1410.3471 [hep-ph]}}.

\bibitem{Azatov:2015oxa}
A.~Azatov, R.~Contino, G.~Panico, and M.~Son, ``{Effective field theory
  analysis of double Higgs boson production via gluon fusion},''
  \href{http://dx.doi.org/10.1103/PhysRevD.92.035001}{{\em Phys. Rev.}
  {\bfseries D92} no.~3, (2015) 035001},
\href{http://arxiv.org/abs/1502.00539}{{\ttfamily arXiv:1502.00539 [hep-ph]}}.

\bibitem{Carvalho:2015ttv}
A.~Carvalho, M.~Dall'Osso, T.~Dorigo, F.~Goertz, C.~A. Gottardo, and M.~Tosi,
  ``{Higgs Pair Production: Choosing Benchmarks With Cluster Analysis},''
  \href{http://dx.doi.org/10.1007/JHEP04(2016)126}{{\em JHEP} {\bfseries 04}
  (2016) 126},
\href{http://arxiv.org/abs/1507.02245}{{\ttfamily arXiv:1507.02245 [hep-ph]}}.

\bibitem{Behr:2015oqq}
J.~K. Behr, D.~Bortoletto, J.~A. Frost, N.~P. Hartland, C.~Issever, and
  J.~Rojo, ``{Boosting Higgs pair production in the $b\bar{b}b\bar{b}$ final
  state with multivariate techniques},''
  \href{http://dx.doi.org/10.1140/epjc/s10052-016-4215-5}{{\em Eur. Phys. J.}
  {\bfseries C76} no.~7, (2016) 386},
\href{http://arxiv.org/abs/1512.08928}{{\ttfamily arXiv:1512.08928 [hep-ph]}}.

\bibitem{Degrassi:2017ucl}
G.~Degrassi, M.~Fedele, and P.~P. Giardino, ``{Constraints on the trilinear
  Higgs self coupling from precision observables},''
\href{http://arxiv.org/abs/1702.01737}{{\ttfamily arXiv:1702.01737 [hep-ph]}}.

\bibitem{Kribs:2017znd}
G.~D. Kribs, A.~Maier, H.~Rzehak, M.~Spannowsky, and P.~Waite, ``{Electroweak
  oblique parameters as a probe of the trilinear Higgs self-interaction},''
\href{http://arxiv.org/abs/1702.07678}{{\ttfamily arXiv:1702.07678 [hep-ph]}}.

\bibitem{McCullough:2013rea}
M.~McCullough, ``{An Indirect Model-Dependent Probe of the Higgs
  Self-Coupling},'' \href{http://dx.doi.org/10.1103/PhysRevD.90.015001,
  10.1103/PhysRevD.92.039903}{{\em Phys. Rev.} {\bfseries D90} no.~1, (2014)
  015001}, \href{http://arxiv.org/abs/1312.3322}{{\ttfamily arXiv:1312.3322
  [hep-ph]}}.
[Erratum: Phys. Rev.D92,no.3,039903(2015)].

\bibitem{Gorbahn:2016uoy}
M.~Gorbahn and U.~Haisch, ``{Indirect probes of the trilinear Higgs coupling:
  $gg \to h$ and $h \to \gamma \gamma$},''
  \href{http://dx.doi.org/10.1007/JHEP10(2016)094}{{\em JHEP} {\bfseries 10}
  (2016) 094},
\href{http://arxiv.org/abs/1607.03773}{{\ttfamily arXiv:1607.03773 [hep-ph]}}.

\bibitem{Degrassi:2016wml}
G.~Degrassi, P.~P. Giardino, F.~Maltoni, and D.~Pagani, ``{Probing the Higgs
  self coupling via single Higgs production at the LHC},''
  \href{http://dx.doi.org/10.1007/JHEP12(2016)080}{{\em JHEP} {\bfseries 12}
  (2016) 080},
\href{http://arxiv.org/abs/1607.04251}{{\ttfamily arXiv:1607.04251 [hep-ph]}}.

\bibitem{Bizon:2016wgr}
W.~Bizon, M.~Gorbahn, U.~Haisch, and G.~Zanderighi, ``{Constraints on the
  trilinear Higgs coupling from vector boson fusion and associated Higgs
  production at the LHC},''
\href{http://arxiv.org/abs/1610.05771}{{\ttfamily arXiv:1610.05771 [hep-ph]}}.

\bibitem{DiVita:2017eyz}
S.~Di~Vita, C.~Grojean, G.~Panico, M.~Riembau, and T.~Vantalon, ``{A global
  view on the Higgs self-coupling},''
\href{http://arxiv.org/abs/1704.01953}{{\ttfamily arXiv:1704.01953 [hep-ph]}}.

\bibitem{Plehn:2005nk}
T.~Plehn and M.~Rauch, ``{The quartic higgs coupling at hadron colliders},''
  \href{http://dx.doi.org/10.1103/PhysRevD.72.053008}{{\em Phys. Rev.}
  {\bfseries D72} (2005) 053008},
\href{http://arxiv.org/abs/hep-ph/0507321}{{\ttfamily arXiv:hep-ph/0507321
  [hep-ph]}}.

\bibitem{Binoth:2006ym}
T.~Binoth, S.~Karg, N.~Kauer, and R.~Ruckl, ``{Multi-Higgs boson production in
  the Standard Model and beyond},''
  \href{http://dx.doi.org/10.1103/PhysRevD.74.113008}{{\em Phys. Rev.}
  {\bfseries D74} (2006) 113008},
\href{http://arxiv.org/abs/hep-ph/0608057}{{\ttfamily arXiv:hep-ph/0608057
  [hep-ph]}}.

\bibitem{Battaglia:2004mw}
{\bfseries CLIC Physics Working Group} Collaboration, E.~Accomando {\em
  et~al.}, ``{Physics at the CLIC multi-TeV linear collider},'' in {\em
  {Proceedings, 11th International Conference on Hadron spectroscopy (Hadron
  2005): Rio de Janeiro, Brazil, August 21-26, 2005}}.
\newblock 2004.
\newblock \href{http://arxiv.org/abs/hep-ph/0412251}{{\ttfamily
  arXiv:hep-ph/0412251 [hep-ph]}}.
\newblock
\url{http://weblib.cern.ch/abstract?CERN-2004-005}.
\newblock

\bibitem{Burgess:2001tj}
C.~P. Burgess, V.~Di~Clemente, and J.~R. Espinosa, ``{Effective operators and
  vacuum instability as heralds of new physics},''
  \href{http://dx.doi.org/10.1088/1126-6708/2002/01/041}{{\em JHEP} {\bfseries
  01} (2002) 041},
\href{http://arxiv.org/abs/hep-ph/0201160}{{\ttfamily arXiv:hep-ph/0201160
  [hep-ph]}}.

\bibitem{Cornwall:1974km}
J.~M. Cornwall, D.~N. Levin, and G.~Tiktopoulos, ``{Derivation of Gauge
  Invariance from High-Energy Unitarity Bounds on the s Matrix},''
  \href{http://dx.doi.org/10.1103/PhysRevD.10.1145,
  10.1103/PhysRevD.11.972}{{\em Phys. Rev.} {\bfseries D10} (1974) 1145}.
[Erratum: Phys. Rev.D11,972(1975)].

\bibitem{Lee:1977yc}
B.~W. Lee, C.~Quigg, and H.~B. Thacker, ``{The Strength of Weak Interactions at
  Very High-Energies and the Higgs Boson Mass},''
\href{http://dx.doi.org/10.1103/PhysRevLett.38.883}{{\em Phys. Rev. Lett.}
  {\bfseries 38} (1977) 883--885}.

\bibitem{Barger:2003rs}
V.~Barger, T.~Han, P.~Langacker, B.~McElrath, and P.~Zerwas, ``{Effects of
  genuine dimension-six Higgs operators},''
  \href{http://dx.doi.org/10.1103/PhysRevD.67.115001}{{\em Phys. Rev.}
  {\bfseries D67} (2003) 115001},
\href{http://arxiv.org/abs/hep-ph/0301097}{{\ttfamily arXiv:hep-ph/0301097
  [hep-ph]}}.

\bibitem{Contino:2013kra}
R.~Contino, M.~Ghezzi, C.~Grojean, M.~M{\"u}hlleitner, and M.~Spira,
  ``{Effective Lagrangian for a light Higgs-like scalar},''
  \href{http://dx.doi.org/10.1007/JHEP07(2013)035}{{\em JHEP} {\bfseries 07}
  (2013) 035},
\href{http://arxiv.org/abs/1303.3876}{{\ttfamily arXiv:1303.3876 [hep-ph]}}.

\bibitem{Pomarol:2014dya}
A.~Pomarol, ``{Higgs Physics},'' in {\em {Proceedings, 2014 European School of
  High-Energy Physics (ESHEP 2014): Garderen, The Netherlands, June 18 - July
  01 2014}}, pp.~59--77.
\newblock 2016.
\newblock \href{http://arxiv.org/abs/1412.4410}{{\ttfamily arXiv:1412.4410
  [hep-ph]}}.
\newblock
\url{http://inspirehep.net/record/1334375/files/arXiv:1412.4410.pdf}.
\newblock

\bibitem{Giudice:2007fh}
G.~F. Giudice, C.~Grojean, A.~Pomarol, and R.~Rattazzi, ``{The
  Strongly-Interacting Light Higgs},''
  \href{http://dx.doi.org/10.1088/1126-6708/2007/06/045}{{\em JHEP} {\bfseries
  06} (2007) 045},
\href{http://arxiv.org/abs/hep-ph/0703164}{{\ttfamily arXiv:hep-ph/0703164
  [hep-ph]}}.

\bibitem{Schuessler:2007av}
A.~Schuessler and D.~Zeppenfeld, ``{Unitarity constraints on MSSM trilinear
  couplings},'' in {\em {SUSY 2007 Proceedings, 15th International Conference
  on Supersymmetry and Unification of Fundamental Interactions, July 26 -
  August 1, 2007, Karlsruhe, Germany}}, pp.~236--239.
\newblock 2007.
\newblock \href{http://arxiv.org/abs/0710.5175}{{\ttfamily arXiv:0710.5175
  [hep-ph]}}.
\newblock
\url{http://www.susy07.uni-karlsruhe.de/Proceedings/proceedings/susy07.pdf}.
\newblock

\bibitem{DiLuzio:2016sur}
L.~Di~Luzio, J.~F. Kamenik, and M.~Nardecchia, ``{Implications of perturbative
  unitarity for scalar di-boson resonance searches at LHC},''
  \href{http://dx.doi.org/10.1140/epjc/s10052-017-4594-2}{{\em Eur. Phys. J.}
  {\bfseries C77} no.~1, (2017) 30},
\href{http://arxiv.org/abs/1604.05746}{{\ttfamily arXiv:1604.05746 [hep-ph]}}.

\bibitem{Lee:1977eg}
B.~W. Lee, C.~Quigg, and H.~B. Thacker, ``{Weak Interactions at Very
  High-Energies: The Role of the Higgs Boson Mass},''
\href{http://dx.doi.org/10.1103/PhysRevD.16.1519}{{\em Phys. Rev.} {\bfseries
  D16} (1977) 1519}.

\bibitem{Luscher:1988gc}
M.~Luscher and P.~Weisz, ``{Is There a Strong Interaction Sector in the
  Standard Lattice Higgs Model?},''
\href{http://dx.doi.org/10.1016/0370-2693(88)91799-6}{{\em Phys. Lett.}
  {\bfseries B212} (1988) 472--478}.

\bibitem{Patel:2015tea}
H.~H. Patel, ``{Package-X: A Mathematica package for the analytic calculation
  of one-loop integrals},''
  \href{http://dx.doi.org/10.1016/j.cpc.2015.08.017}{{\em Comput. Phys.
  Commun.} {\bfseries 197} (2015) 276--290},
\href{http://arxiv.org/abs/1503.01469}{{\ttfamily arXiv:1503.01469 [hep-ph]}}.

\bibitem{Goertz:2015nkp}
F.~Goertz, J.~F. Kamenik, A.~Katz, and M.~Nardecchia, ``{Indirect Constraints
  on the Scalar Di-Photon Resonance at the LHC},''
  \href{http://dx.doi.org/10.1007/JHEP05(2016)187}{{\em JHEP} {\bfseries 05}
  (2016) 187},
\href{http://arxiv.org/abs/1512.08500}{{\ttfamily arXiv:1512.08500 [hep-ph]}}.

\bibitem{Dolan:2012ac}
M.~J. Dolan, C.~Englert, and M.~Spannowsky, ``{New Physics in LHC Higgs boson
  pair production},'' \href{http://dx.doi.org/10.1103/PhysRevD.87.055002}{{\em
  Phys. Rev.} {\bfseries D87} no.~5, (2013) 055002},
\href{http://arxiv.org/abs/1210.8166}{{\ttfamily arXiv:1210.8166 [hep-ph]}}.

\bibitem{No:2013wsa}
J.~M. No and M.~Ramsey-Musolf, ``{Probing the Higgs Portal at the LHC Through
  Resonant di-Higgs Production},''
  \href{http://dx.doi.org/10.1103/PhysRevD.89.095031}{{\em Phys. Rev.}
  {\bfseries D89} no.~9, (2014) 095031},
\href{http://arxiv.org/abs/1310.6035}{{\ttfamily arXiv:1310.6035 [hep-ph]}}.

\bibitem{Chen:2014ask}
C.-Y. Chen, S.~Dawson, and I.~M. Lewis, ``{Exploring resonant di-Higgs boson
  production in the Higgs singlet model},''
  \href{http://dx.doi.org/10.1103/PhysRevD.91.035015}{{\em Phys. Rev.}
  {\bfseries D91} no.~3, (2015) 035015},
\href{http://arxiv.org/abs/1410.5488}{{\ttfamily arXiv:1410.5488 [hep-ph]}}.

\bibitem{Martin-Lozano:2015dja}
V.~Mart{\'i}n~Lozano, J.~M. Moreno, and C.~B. Park, ``{Resonant Higgs boson
  pair production in the $ hh\to b\overline{b}\ WW\to
  b\overline{b}{\ell}^{+}\nu {\ell}^{-}\overline{\nu} $ decay channel},''
  \href{http://dx.doi.org/10.1007/JHEP08(2015)004}{{\em JHEP} {\bfseries 08}
  (2015) 004},
\href{http://arxiv.org/abs/1501.03799}{{\ttfamily arXiv:1501.03799 [hep-ph]}}.

\bibitem{Godunov:2015nea}
S.~I. Godunov, A.~N. Rozanov, M.~I. Vysotsky, and E.~V. Zhemchugov,
  ``{Extending the Higgs sector: an extra singlet},''
  \href{http://dx.doi.org/10.1140/epjc/s10052-015-3826-6}{{\em Eur. Phys. J.}
  {\bfseries C76} (2016) 1},
\href{http://arxiv.org/abs/1503.01618}{{\ttfamily arXiv:1503.01618 [hep-ph]}}.

\bibitem{Banerjee:2016nzb}
S.~Banerjee, B.~Batell, and M.~Spannowsky, ``{Invisible decays in Higgs boson
  pair production},'' \href{http://dx.doi.org/10.1103/PhysRevD.95.035009}{{\em
  Phys. Rev.} {\bfseries D95} no.~3, (2017) 035009},
\href{http://arxiv.org/abs/1608.08601}{{\ttfamily arXiv:1608.08601 [hep-ph]}}.

\bibitem{Huang:2017jws}
T.~Huang, J.~M. No, L.~Perni{\'e}, M.~Ramsey-Musolf, A.~Safonov, M.~Spannowsky,
  and P.~Winslow, ``{Resonant Di-Higgs Production in the $b{\bar b}WW$ Channel:
  Probing the Electroweak Phase Transition at the LHC},''
\href{http://arxiv.org/abs/1701.04442}{{\ttfamily arXiv:1701.04442 [hep-ph]}}.

\bibitem{Lewis:2017dme}
I.~M. Lewis and M.~Sullivan, ``{Benchmarks for the Singlet Extended Standard
  Model at the LHC},''
\href{http://arxiv.org/abs/1701.08774}{{\ttfamily arXiv:1701.08774 [hep-ph]}}.

\bibitem{Dawson:2015oha}
S.~Dawson, A.~Ismail, and I.~Low, ``{What's in the loop? The anatomy of double
  Higgs production},'' \href{http://dx.doi.org/10.1103/PhysRevD.91.115008}{{\em
  Phys. Rev.} {\bfseries D91} no.~11, (2015) 115008},
\href{http://arxiv.org/abs/1504.05596}{{\ttfamily arXiv:1504.05596 [hep-ph]}}.

\bibitem{Grober:2016wmf}
R.~Gr{\"o}ber, M.~M{\"u}hlleitner, and M.~Spira, ``{Signs of Composite Higgs
  Pair Production at Next-to-Leading Order},''
  \href{http://dx.doi.org/10.1007/JHEP06(2016)080}{{\em JHEP} {\bfseries 06}
  (2016) 080},
\href{http://arxiv.org/abs/1602.05851}{{\ttfamily arXiv:1602.05851 [hep-ph]}}.

\bibitem{Agostini:2016vze}
A.~Agostini, G.~Degrassi, R.~Gr{\"o}ber, and P.~Slavich, ``{NLO-QCD corrections
  to Higgs pair production in the MSSM},''
  \href{http://dx.doi.org/10.1007/JHEP04(2016)106}{{\em JHEP} {\bfseries 04}
  (2016) 106},
\href{http://arxiv.org/abs/1601.03671}{{\ttfamily arXiv:1601.03671 [hep-ph]}}.

\bibitem{delAguila:2010mx}
F.~del Aguila, J.~de~Blas, and M.~Perez-Victoria, ``{Electroweak Limits on
  General New Vector Bosons},''
  \href{http://dx.doi.org/10.1007/JHEP09(2010)033}{{\em JHEP} {\bfseries 09}
  (2010) 033},
\href{http://arxiv.org/abs/1005.3998}{{\ttfamily arXiv:1005.3998 [hep-ph]}}.

\bibitem{Biggio:2016wyy}
C.~Biggio, M.~Bordone, L.~Di~Luzio, and G.~Ridolfi, ``{Massive vectors and loop
  observables: the $g-2$ case},''
  \href{http://dx.doi.org/10.1007/JHEP10(2016)002}{{\em JHEP} {\bfseries 10}
  (2016) 002},
\href{http://arxiv.org/abs/1607.07621}{{\ttfamily arXiv:1607.07621 [hep-ph]}}.

\bibitem{deBlas:2014mba}
J.~de~Blas, M.~Chala, M.~Perez-Victoria, and J.~Santiago, ``{Observable Effects
  of General New Scalar Particles},''
  \href{http://dx.doi.org/10.1007/JHEP04(2015)078}{{\em JHEP} {\bfseries 04}
  (2015) 078},
\href{http://arxiv.org/abs/1412.8480}{{\ttfamily arXiv:1412.8480 [hep-ph]}}.

\bibitem{DiLuzio:2015oha}
L.~Di~Luzio, R.~Gr{\" o}ber, J.~F. Kamenik, and M.~Nardecchia, ``{Accidental
  matter at the LHC},'' \href{http://dx.doi.org/10.1007/JHEP07(2015)074}{{\em
  JHEP} {\bfseries 07} (2015) 074},
\href{http://arxiv.org/abs/1504.00359}{{\ttfamily arXiv:1504.00359 [hep-ph]}}.

\bibitem{Jiang:2016czg}
Y.~Jiang and M.~Trott, ``{On the non-minimal character of the SMEFT},''
\href{http://arxiv.org/abs/1612.02040}{{\ttfamily arXiv:1612.02040 [hep-ph]}}.

\bibitem{Gupta:2013zza}
R.~S. Gupta, H.~Rzehak, and J.~D. Wells, ``{How well do we need to measure the
  Higgs boson mass and self-coupling?},''
  \href{http://dx.doi.org/10.1103/PhysRevD.88.055024}{{\em Phys. Rev.}
  {\bfseries D88} (2013) 055024},
\href{http://arxiv.org/abs/1305.6397}{{\ttfamily arXiv:1305.6397 [hep-ph]}}.

\bibitem{Efrati:2014uta}
A.~Efrati and Y.~Nir, ``{What if $\lambda_{hhh}\neq 3m_h^2/v$},''
\href{http://arxiv.org/abs/1401.0935}{{\ttfamily arXiv:1401.0935 [hep-ph]}}.

\bibitem{Lopez-Val:2014jva}
D.~L{\'o}pez-Val and T.~Robens, ``{$\Delta r$ and the W-boson mass in the
  singlet extension of the standard model},''
  \href{http://dx.doi.org/10.1103/PhysRevD.90.114018}{{\em Phys. Rev.}
  {\bfseries D90} (2014) 114018},
\href{http://arxiv.org/abs/1406.1043}{{\ttfamily arXiv:1406.1043 [hep-ph]}}.

\bibitem{Aad:2015pla}
{\bfseries ATLAS} Collaboration, G.~Aad {\em et~al.}, ``{Constraints on new
  phenomena via Higgs boson couplings and invisible decays with the ATLAS
  detector},'' \href{http://dx.doi.org/10.1007/JHEP11(2015)206}{{\em JHEP}
  {\bfseries 11} (2015) 206},
\href{http://arxiv.org/abs/1509.00672}{{\ttfamily arXiv:1509.00672 [hep-ex]}}.

\bibitem{Robens:2016xkb}
T.~Robens and T.~Stefaniak, ``{LHC Benchmark Scenarios for the Real Higgs
  Singlet Extension of the Standard Model},''
  \href{http://dx.doi.org/10.1140/epjc/s10052-016-4115-8}{{\em Eur. Phys. J.}
  {\bfseries C76} no.~5, (2016) 268},
\href{http://arxiv.org/abs/1601.07880}{{\ttfamily arXiv:1601.07880 [hep-ph]}}.

\bibitem{Espinosa:2011ax}
J.~R. Espinosa, T.~Konstandin, and F.~Riva, ``{Strong Electroweak Phase
  Transitions in the Standard Model with a Singlet},''
  \href{http://dx.doi.org/10.1016/j.nuclphysb.2011.09.010}{{\em Nucl. Phys.}
  {\bfseries B854} (2012) 592--630},
\href{http://arxiv.org/abs/1107.5441}{{\ttfamily arXiv:1107.5441 [hep-ph]}}.

\bibitem{Camargo-Molina:2013qva}
J.~E. Camargo-Molina, B.~O'Leary, W.~Porod, and F.~Staub,
  ``{$\mathbf{Vevacious}$: A Tool For Finding The Global Minima Of One-Loop
  Effective Potentials With Many Scalars},''
  \href{http://dx.doi.org/10.1140/epjc/s10052-013-2588-2}{{\em Eur. Phys. J.}
  {\bfseries C73} no.~10, (2013) 2588},
\href{http://arxiv.org/abs/1307.1477}{{\ttfamily arXiv:1307.1477 [hep-ph]}}.

\bibitem{Camargo-Molina:2014pwa}
J.~E. Camargo-Molina, B.~Garbrecht, B.~O'Leary, W.~Porod, and F.~Staub,
  ``{Constraining the Natural MSSM through tunneling to color-breaking vacua at
  zero and non-zero temperature},''
  \href{http://dx.doi.org/10.1016/j.physletb.2014.08.036}{{\em Phys. Lett.}
  {\bfseries B737} (2014) 156--161},
\href{http://arxiv.org/abs/1405.7376}{{\ttfamily arXiv:1405.7376 [hep-ph]}}.

\bibitem{Staub:2008uz}
F.~Staub, ``{SARAH},''
\href{http://arxiv.org/abs/0806.0538}{{\ttfamily arXiv:0806.0538 [hep-ph]}}.

\bibitem{Staub:2009bi}
F.~Staub, ``{From Superpotential to Model Files for FeynArts and
  CalcHep/CompHep},'' \href{http://dx.doi.org/10.1016/j.cpc.2010.01.011}{{\em
  Comput. Phys. Commun.} {\bfseries 181} (2010) 1077--1086},
\href{http://arxiv.org/abs/0909.2863}{{\ttfamily arXiv:0909.2863 [hep-ph]}}.

\bibitem{Staub:2010jh}
F.~Staub, ``{Automatic Calculation of supersymmetric Renormalization Group
  Equations and Self Energies},''
  \href{http://dx.doi.org/10.1016/j.cpc.2010.11.030}{{\em Comput. Phys.
  Commun.} {\bfseries 182} (2011) 808--833},
\href{http://arxiv.org/abs/1002.0840}{{\ttfamily arXiv:1002.0840 [hep-ph]}}.

\bibitem{Staub:2012pb}
F.~Staub, ``{SARAH 3.2: Dirac Gauginos, UFO output, and more},''
  \href{http://dx.doi.org/10.1016/j.cpc.2013.02.019}{{\em Comput. Phys.
  Commun.} {\bfseries 184} (2013) 1792--1809},
\href{http://arxiv.org/abs/1207.0906}{{\ttfamily arXiv:1207.0906 [hep-ph]}}.

\bibitem{Staub:2013tta}
F.~Staub, ``{SARAH 4 : A tool for (not only SUSY) model builders},''
  \href{http://dx.doi.org/10.1016/j.cpc.2014.02.018}{{\em Comput. Phys.
  Commun.} {\bfseries 185} (2014) 1773--1790},
\href{http://arxiv.org/abs/1309.7223}{{\ttfamily arXiv:1309.7223 [hep-ph]}}.

\bibitem{Buttazzo:2015bka}
D.~Buttazzo, F.~Sala, and A.~Tesi, ``{Singlet-like Higgs bosons at present and
  future colliders},'' \href{http://dx.doi.org/10.1007/JHEP11(2015)158}{{\em
  JHEP} {\bfseries 11} (2015) 158},
\href{http://arxiv.org/abs/1505.05488}{{\ttfamily arXiv:1505.05488 [hep-ph]}}.

\bibitem{Kanemura:2016lkz}
S.~Kanemura, M.~Kikuchi, and K.~Yagyu, ``{One-loop corrections to the Higgs
  self-couplings in the singlet extension},''
  \href{http://dx.doi.org/10.1016/j.nuclphysb.2017.02.004}{{\em Nucl. Phys.}
  {\bfseries B917} (2017) 154--177},
\href{http://arxiv.org/abs/1608.01582}{{\ttfamily arXiv:1608.01582 [hep-ph]}}.

\bibitem{Camargo-Molina:2016moz}
J.~E. Camargo-Molina, A.~P. Morais, R.~Pasechnik, M.~O.~P. Sampaio, and
  J.~Wess{\'e}n, ``{All one-loop scalar vertices in the effective potential
  approach},'' \href{http://dx.doi.org/10.1007/JHEP08(2016)073}{{\em JHEP}
  {\bfseries 08} (2016) 073},
\href{http://arxiv.org/abs/1606.07069}{{\ttfamily arXiv:1606.07069 [hep-ph]}}.

\bibitem{Kanemura:2004mg}
S.~Kanemura, Y.~Okada, E.~Senaha, and C.~P. Yuan, ``{Higgs coupling constants
  as a probe of new physics},''
  \href{http://dx.doi.org/10.1103/PhysRevD.70.115002}{{\em Phys. Rev.}
  {\bfseries D70} (2004) 115002},
\href{http://arxiv.org/abs/hep-ph/0408364}{{\ttfamily arXiv:hep-ph/0408364
  [hep-ph]}}.

\bibitem{Bernon:2015wef}
J.~Bernon, J.~F. Gunion, H.~E. Haber, Y.~Jiang, and S.~Kraml, ``{Scrutinizing
  the alignment limit in two-Higgs-doublet models. II. m$_H$=125 GeV},''
  \href{http://dx.doi.org/10.1103/PhysRevD.93.035027}{{\em Phys. Rev.}
  {\bfseries D93} no.~3, (2016) 035027},
\href{http://arxiv.org/abs/1511.03682}{{\ttfamily arXiv:1511.03682 [hep-ph]}}.

\bibitem{Baglio:2014nea}
J.~Baglio, O.~Eberhardt, U.~Nierste, and M.~Wiebusch, ``{Benchmarks for Higgs
  Pair Production and Heavy Higgs boson Searches in the Two-Higgs-Doublet Model
  of Type II},'' \href{http://dx.doi.org/10.1103/PhysRevD.90.015008}{{\em Phys.
  Rev.} {\bfseries D90} no.~1, (2014) 015008},
\href{http://arxiv.org/abs/1403.1264}{{\ttfamily arXiv:1403.1264 [hep-ph]}}.

\bibitem{Ginzburg:2015yva}
I.~F. Ginzburg, ``{Triple Higgs coupling in the most general 2HDM at SM-like
  scenario},'' \href{http://dx.doi.org/10.1140/epjc/s10052-016-4559-x}{{\em
  Eur. Phys. J.} {\bfseries C77} no.~1, (2017) 9},
\href{http://arxiv.org/abs/1510.08270}{{\ttfamily arXiv:1510.08270 [hep-ph]}}.

\bibitem{Chen:2006pb}
M.-C. Chen, S.~Dawson, and T.~Krupovnickas, ``{Higgs triplets and limits from
  precision measurements},''
  \href{http://dx.doi.org/10.1103/PhysRevD.74.035001}{{\em Phys. Rev.}
  {\bfseries D74} (2006) 035001},
\href{http://arxiv.org/abs/hep-ph/0604102}{{\ttfamily arXiv:hep-ph/0604102
  [hep-ph]}}.

\bibitem{Lynn:1990zk}
B.~W. Lynn and E.~Nardi, ``{Radiative corrections in unconstrained SU(2) x U(1)
  and the top mass problem},''
\href{http://dx.doi.org/10.1016/0550-3213(92)90486-U}{{\em Nucl. Phys.}
  {\bfseries B381} (1992) 467--500}.

\bibitem{Blank:1997qa}
T.~Blank and W.~Hollik, ``{Precision observables in SU(2) x U(1) models with an
  additional Higgs triplet},''
  \href{http://dx.doi.org/10.1016/S0550-3213(97)00785-2}{{\em Nucl. Phys.}
  {\bfseries B514} (1998) 113--134},
\href{http://arxiv.org/abs/hep-ph/9703392}{{\ttfamily arXiv:hep-ph/9703392
  [hep-ph]}}.

\bibitem{Olive:2016xmw}
{\bfseries Particle Data Group} Collaboration, C.~Patrignani {\em et~al.},
  ``{Review of Particle Physics},''
\href{http://dx.doi.org/10.1088/1674-1137/40/10/100001}{{\em Chin. Phys.}
  {\bfseries C40} no.~10, (2016) 100001}.

\bibitem{arXiv:0811.4169}
P.~Bechtle, O.~Brein, S.~Heinemeyer, G.~Weiglein, and K.~E. Williams,
  ``{HiggsBounds: Confronting Arbitrary Higgs Sectors with Exclusion Bounds
  from LEP and the Tevatron},''
  \href{http://dx.doi.org/10.1016/j.cpc.2009.09.003}{{\em Comput. Phys.
  Commun.} {\bfseries 181} (2010) 138--167},
\href{http://arxiv.org/abs/0811.4169}{{\ttfamily arXiv:0811.4169 [hep-ph]}}.

\bibitem{arXiv:1311.0055}
P.~Bechtle {\em et~al.}, ``{HiggsBounds-4: Improved Tests of Extended Higgs
  Sectors against Exclusion Bounds from LEP, the Tevatron and the LHC},'' {\em
  Eur. Phys. J.} {\bfseries C74} (2014) 2693,
\href{http://arxiv.org/abs/1311.0055}{{\ttfamily arXiv:1311.0055 [hep-ph]}}.

\bibitem{arXiv:1507.06706}
P.~Bechtle, S.~Heinemeyer, O.~Stal, T.~Stefaniak, and G.~Weiglein, ``{Applying
  Exclusion Likelihoods from LHC Searches to Extended Higgs Sectors},''
\href{http://arxiv.org/abs/1507.06706}{{\ttfamily arXiv:1507.06706 [hep-ph]}}.

\bibitem{Khan:2016sxm}
N.~Khan, ``{Exploring Hyperchargeless Higgs Triplet Model up to the Planck
  Scale},''
\href{http://arxiv.org/abs/1610.03178}{{\ttfamily arXiv:1610.03178 [hep-ph]}}.

\bibitem{Baglio:2016ijw}
J.~Baglio and C.~Weiland, ``{Heavy neutrino impact on the triple Higgs
  coupling},'' \href{http://dx.doi.org/10.1103/PhysRevD.94.013002}{{\em Phys.
  Rev.} {\bfseries D94} no.~1, (2016) 013002},
\href{http://arxiv.org/abs/1603.00879}{{\ttfamily arXiv:1603.00879 [hep-ph]}}.

\bibitem{Baglio:2016bop}
J.~Baglio and C.~Weiland, ``{The triple Higgs coupling: A new probe of
  low-scale seesaw models},''
\href{http://arxiv.org/abs/1612.06403}{{\ttfamily arXiv:1612.06403 [hep-ph]}}.

\bibitem{Rose:2015fua}
L.~Delle~Rose, C.~Marzo, and A.~Urbano, ``{On the stability of the electroweak
  vacuum in the presence of low-scale seesaw models},''
  \href{http://dx.doi.org/10.1007/JHEP12(2015)050}{{\em JHEP} {\bfseries 12}
  (2015) 050},
\href{http://arxiv.org/abs/1506.03360}{{\ttfamily arXiv:1506.03360 [hep-ph]}}.

\bibitem{Isidori:2001bm}
G.~Isidori, G.~Ridolfi, and A.~Strumia, ``{On the metastability of the standard
  model vacuum},'' \href{http://dx.doi.org/10.1016/S0550-3213(01)00302-9}{{\em
  Nucl. Phys.} {\bfseries B609} (2001) 387--409},
\href{http://arxiv.org/abs/hep-ph/0104016}{{\ttfamily arXiv:hep-ph/0104016
  [hep-ph]}}.

\bibitem{DiLuzio:2015iua}
L.~Di~Luzio, G.~Isidori, and G.~Ridolfi, ``{Stability of the electroweak ground
  state in the Standard Model and its extensions},''
  \href{http://dx.doi.org/10.1016/j.physletb.2015.12.009}{{\em Phys. Lett.}
  {\bfseries B753} (2016) 150--160},
\href{http://arxiv.org/abs/1509.05028}{{\ttfamily arXiv:1509.05028 [hep-ph]}}.

\end{thebibliography}\endgroup

\end{document}